\documentclass[12pt,titlepage]{article}%
\usepackage{graphicx}
\usepackage{amsmath}
\usepackage{amsfonts}
\usepackage{amssymb}
\usepackage{color}%
\providecommand{\U}[1]{\protect\rule{.1in}{.1in}}
\begin{document}

\title{Valuation and Hedging of the \\Ruin-Contingent Life Annuity (RCLA)}
\author{H. Huang, M. A. Milevsky\thanks{Huang is Professor of Mathematics and
Statistics at York University. Milevsky is Associate Professor of Finance,
York University, and Executive Director of the IFID Centre. Salisbury is
Professor of Mathematics and Statistics at York University, all in Toronto,
Canada. The contact author (Milevsky) can be reached via email at:
milevsky@yorku.ca. The authors acknowledge the helpful comments of seminar
participants at the Department of Risk Management and Insurance at The Wharton
School, as well as seminar participants at Monash University, Melbourne, The
University of New South Wales and the University of Technology, Sydney. In
particular the authors would like to acknowledge helpful comments from Carl
Chiarella, Neil Doherty, Olivia Mitchell, and Eckhard Platen. Huang's and
Salisbury's research is supported in part by NSERC and MITACS. } and T.S. Salisbury}
\date{Version: 16 May 2012}

\maketitle

\begin{abstract} 

\emph{VALUATION AND HEDGING OF THE RUIN-CONTINGENT LIFE ANNUITY}\medskip

This paper analyzes a novel type of mortality contingent-claim called a
ruin-contingent life annuity (RCLA). This product fuses together a
path-dependent equity put option with a ``personal longevity'' call option. The
annuitant's (i.e. long position) payoff from a generic RCLA is \$1 of income
per year for life, akin to a defined benefit pension, but deferred until a
pre-specified financial diffusion process hits zero. We derive the PDE and
relevant boundary conditions satisfied by the RCLA value (i.e. the hedging
cost) assuming a complete market where No Arbitrage is possible. We then
describe some efficient numerical techniques and provide estimates of a
typical RCLA under a variety of realistic parameters. 

The motivation for studying the RCLA on a stand-alone basis is two-fold. First, it is implicitly embedded in approximately \$1 trillion worth of U.S. variable annuity (VA) policies; which have recently attracted scrutiny from financial analysts and regulators. Second, the U.S. administration -- both Treasury and Department of Labor -- have been encouraging Defined Contribution (401k) plans to offer stand-alone longevity insurance to participants, and we believe the RCLA would be an ideal and cost effective candidate for that job.

\end{abstract}

\section{Introduction}

Among the expanding universe of derivative securities priced off non-financial 
state variables, a recent innovation has been the
mortality-contingent claim. As its name suggests, a mortality-contingent claim
is a derivative product whose payoff is dependent or linked to the mortality
status of an underlying reference life or pool of lives. The simplest and
perhaps the most trivial mortality-contingent claim is a personal life
insurance policy with a face value of one million dollars for example. In this
case, the underlying state variable is the (binary) life status of the
insured. If and when it jumps from the value of one (alive) to the value of
zero (dead) the beneficiary of the life insurance policy receives a payout of
one million dollars. Another equally trivial example is a life or pension
annuity policy which provides monthly income until the annuitant dies. Payment
for these options can be made up-front, as in the case of a pension income
annuity, or by installments as in the case of a life insurance policy. Indeed,
the analogy to credit default swaps is obvious and it is said that much of the
technology -- such as Gaussian copulas and reduced form hazard rate models --
which are (rightfully or wrongfully) used for pricing credit derivatives can
be traced to the actuarial science behind the pricing of insurance claims.

Yet, in the past these pure mortality-contingent claims have been (perhaps
rightfully) ignored\footnote{There are some exceptions, for example the 2006
article in the Journal of Derivatives by Stone and Zissu on the topic of
securitizing life insurance settlements.} by the mainstream quant community
primarily because of the law of large numbers. It dictates that a large-enough
portfolio of policies held by a large insurance company should diversify away
all risk. Under this theory pricing collapsed to rather trivial
time-value-of-money calculations based on cash-flows that are highly
predictable in aggregate.

However this conventional viewpoint came into question when, in the early part
of this decade, a number of large insurance companies began offering equity
put options with rather complex optionality that was directly tied to the
mortality status of the insured. These variable annuity (VA) policies, as they
are commonly known, have been the source of much public and regulatory
consternation in late 2008 and early 2009, as the required insurance reserves
mushroomed.  An additional source of interest, not directly
addressed in this paper, is the emergence of actuarial evidence that mortality
itself contains a stochastic component. 
See, for example, Dawson, Dowd, Cairns and Blake (2010), or Schulze, Post (2010).

Motivated by all of this, in this paper we value and provide hedging guidance
on a type of product called a ruin-contingent life annuity (RCLA). The RCLA
provides the buyer with a type of insurance against the joint occurrence of
two separate (and likely independent) events; the two events are \emph{under
average} investment returns and \emph{above average} longevity. The RCLA
behaves like a pension annuity that provides lifetime income, but only in
\emph{bad} \emph{economic} scenarios. In the good scenarios, properly defined,
it pays nothing. The RCLA is obviously (much) cheaper than a generic life
annuity which provides income under all economic scenarios. We will argue that
the RCLA is a fundamental mortality-contingent building block of all VA
\textquotedblleft income guarantees\textquotedblright\ in the sense that it is
not muddled by tax-frictions and other institutional issues. At the same time
it retains many of the real-world features embedded within these policies. At
the very least this article should provide an introduction to what we label
\textsl{finsurance} -- products that combine financial and insurance options
in one package.

Research into longevity insurance and life annuities in general, has increased in prominence and intensity – especially within the scholarly literature -- during the last decade or so. Indeed, there is a growing awareness that most individuals are endowed with some form of longevity insurance – in the form of government social security – and must figure out how to optimize its usage. See, for example, Sun and Webb (2011) for a recent discussion of this within the content of delaying Social Security. Researchers are trying to develop a better understand how other assets might reduce the demand for longevity insurance, see for example Davidoff (2009). Many countries are struggling with the question of how to properly design  a life annuity market. See for example Fong, Mitchell and Koh (2011). In this paper we take a slightly different approach and discuss product innovation.

In a recent article, Scott, Watson and Hu (2011) discussed the characteristics that make-up an ideal (or better) annuity. Using microeconomic welfare analysis, they concluded that innovation in the field should focus on developing products that add survival contingencies to assets commonly held by individuals in retirement. Our current paper is along the same lines in that we actually construct and actually price such a product.

Huang, Milevsky and Salisbury (2009) motivated the need for a stand-alone
ruin-contingent life annuity (RCLA), albeit without deriving any valuation
relationships. Practitioners and regulators have gone on to discuss the framework
for offering such products (motivated in part by the above article), under such
names as {\it contingent deferred annuities} or {\it hybrid annuities} -- see for example
Festa (2012). 
In this article we provide the valuation and hedging machinery
for the RCLA, in a complete market setting (i.e. assuming no arbitrage). In
terms of its position within the actuarial and finance literature, the RCLA is
effectively a type of annuity option, and so this work is related to Ballotta
and Haberman (2003), Deelstra, Vanmaele and Vyncke (2010), as well as 
Hardy (2003) or Boyle and Hardy (2003) in which similar complete market techniques 
are relied upon. In a subsequent paper we plan to
describe the impact of incomplete markets and other frictions.


\subsection{How Does the RCLA Work?}

The RCLA is based on a reference portfolio index (RPI), a.k.a. the state
variable, upon which the income/pension annuity start-date is based. The RPI
is initiated at an artificial level of \$100, for example, and consists of a
broad portfolio of stocks (for example the SP500 or Russell 3000 Index).
However at the end of each day, week or month the RPI is adjusted for total
returns (plus or minus) and by a fixed cash outflow (minus) that reduces the
RPI. The cash outflow can be constant in nominal terms or constant in real
terms or something in between. The income annuity embedded within the RCLA
begins payments if-and-when the RPI hits zero. Figure \#1 provides an example
of a possible sample-path for the RPI in discrete time.%

\begin{figure}
\begin{center}
\includegraphics[scale=.5]{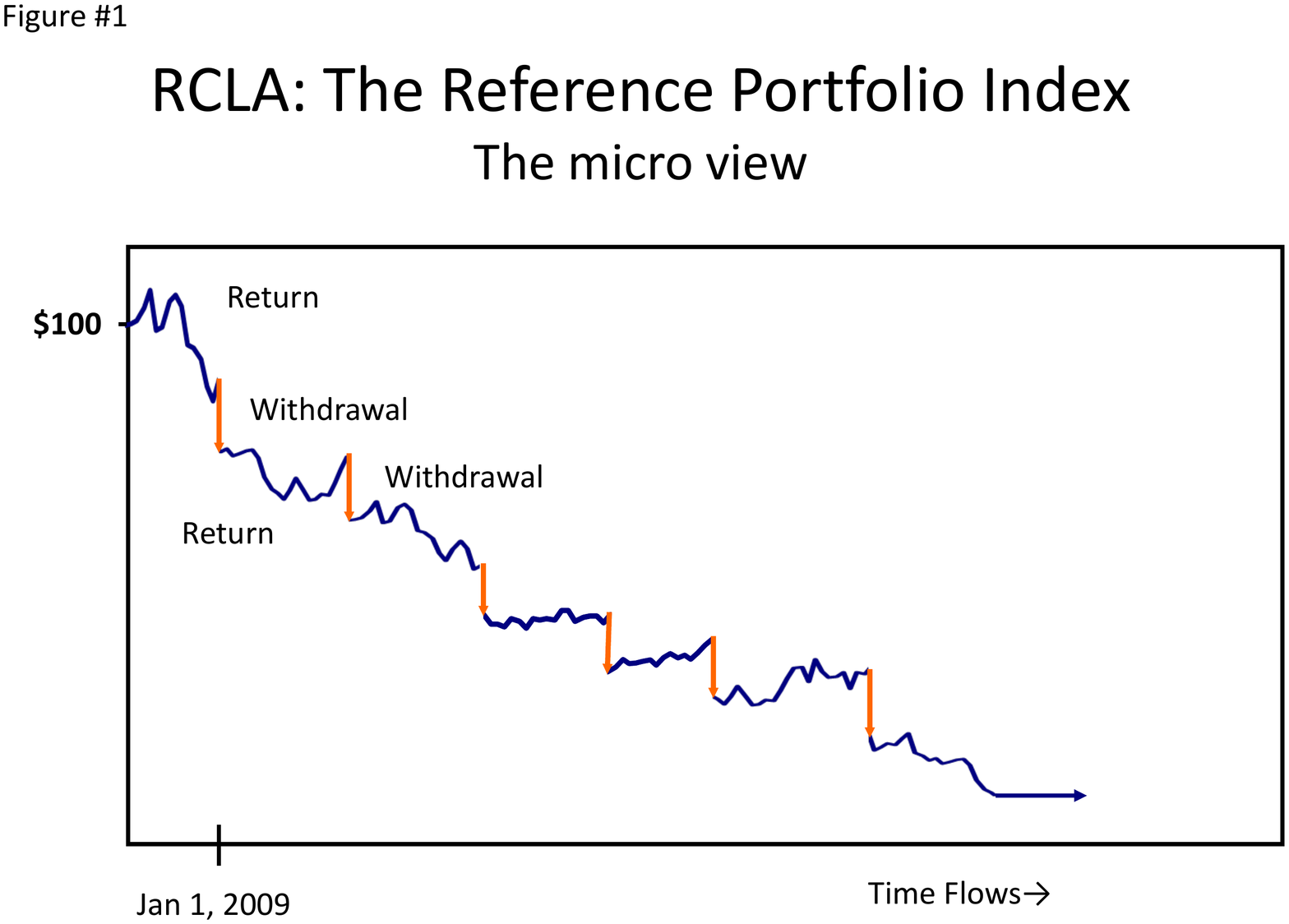} 
\label{figure1}
\end{center}
\end{figure}

Here is a detailed example that should help explain the mechanics of the RPI
and the stochastic annuity start date. Assume that the Russell 3000 index is
at a level of \$100 on January 1st, 2009. 
If under a pre-specified withdrawal
rate of \$7 we assume that during January 2009 the Russell 3000 total return
was a nominal 2\%, then the level of a vintage 2009 RPI on the first day of
February 2009 would be $\$100(1.02)-(7/12)=\$101.\,\allowbreak42$. The annual
withdrawal rate of \$7 is divided by 12 to create the monthly withdrawal,
which can also be adjusted for inflation. The same calculation algorithm
continues each month. Think of the RPI as mimicking the behavior of a
retirement drawdown portfolio.

Now, if and when this (vintage) 2009 RPI ever hits zero, the insurance company
would then commence making \$1 for life payments (either nominal or
inflation-adjusted) to the annuitant who bought the product in January 2009,
as long as they were still alive. Figure \#2 graphically illustrates how the
performance of the RPI would trigger the lifetime income payment. Under path
\#1 in which the RPI hits zero twenty years after purchase, the income would
start at the age of 80. Under path \#2 where the RPI never hits zero, the
annuitant would receive nothing from the insurance company.

\begin{figure}
\begin{center}
\includegraphics[scale=.5]{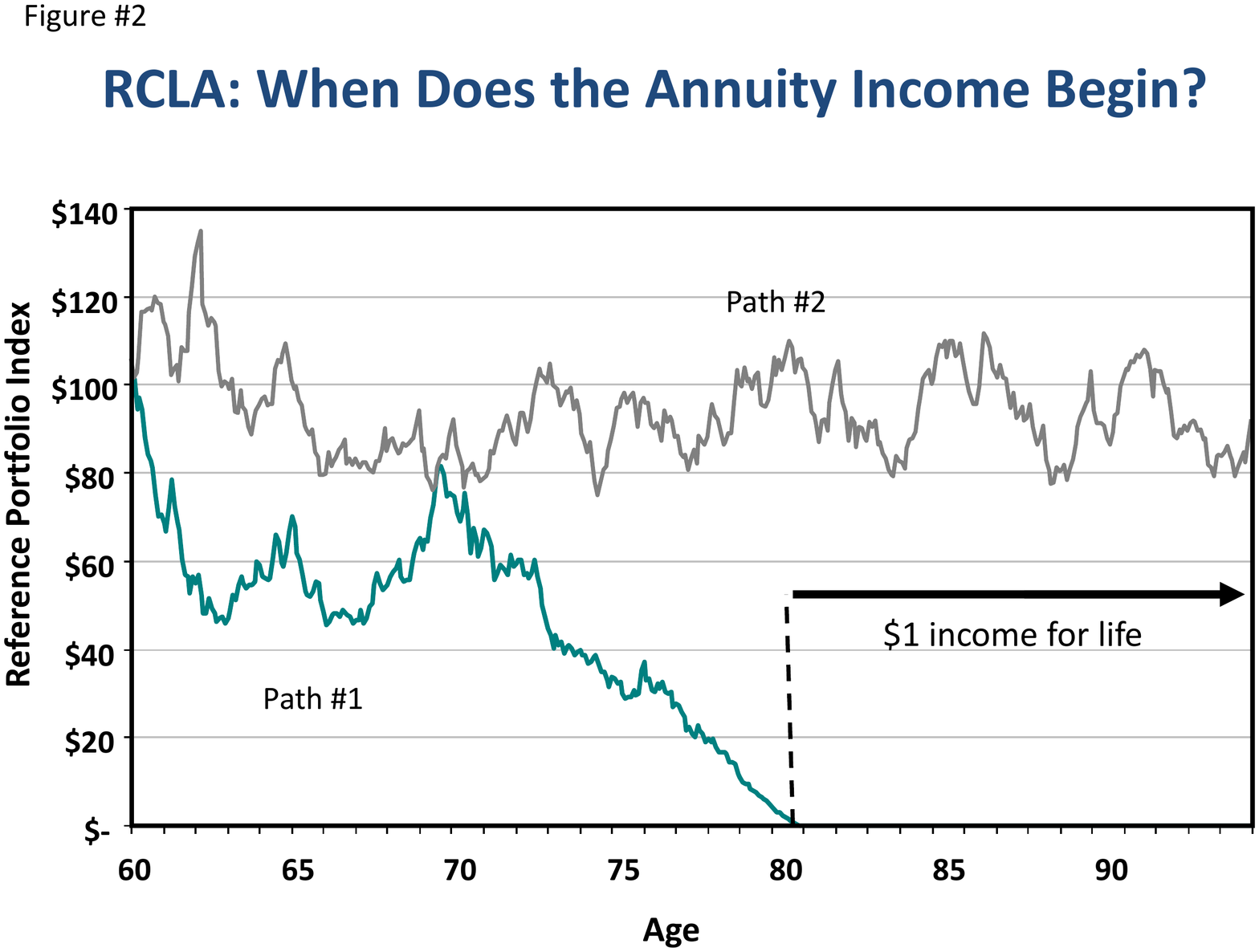} 
\label{figure2}
\end{center}
\end{figure}

A generic RCLA is defined in units of \$1, so if the annuitant purchased 7
units, they would continue to receive the same \$7 of income without any
disruptions to their standard of living. At inception the retiree buying the
RCLA could select from a range of withdrawal rates, for example 5\%, 6\% or
7\%, assuming the insurance company was willing to offer a menu of spending
rates (at different prices, of course.) Likewise, the annuitant could specify
nominal payments of \$1 for life or real payments of \$1 for life, which would
obviously impact pricing as well.

To be precise, and when necessary, we will use the notation 
$W_t=W_t(z,\gamma)$
to denote the level/value of the reference portfolio index in year
$(z+t)$, where the initial withdrawal rate in year $z$, was set at $\gamma$
percent of the initial value $I_{0}$. In other words, $W_t$ is the state variable 
underlying the derivative's payout function.

It is worth pointing out that, from the point of view of the insurance company
offering an RCLA, this is a complete-markets product, that can be perfectly
hedged (at least under our assumptions). Thus the price or value we will
compute below is really measuring the company's hedging cost. This may differ
from the economic value an individual client places on the product, since from
the client's point of view, the market is incomplete and mortality risk is
unhedgeable. What makes a hedge possible for the company is the law of large
numbers -- after selling many individual contracts, the total cash flows due
to mortality become essentially deterministic, leaving only cash flows due to
market fluctuations to be hedged. We will comment further on this issue below.

\subsection{Agenda for the Paper}

In section \#2 we briefly review the pricing of generic life annuities, which
also helps introduce notation and provides some basic intuition for the RCLA.
Section \#3 formally introduces the concept of \textquotedblleft
ruin\textquotedblright\ under the relevant diffusion process, which becomes
the trigger for the RCLA. Section \#4, which is the core of the paper,
introduces, values, and then describes the hedge for a basic RCLA. Section \#5
describes some advanced products in which the payoff and ruin-trigger are
non-constant. It also discusses the connection between RCLA values and the
popular {\em Guaranteed Living Withdrawal Benefits} (GLWB) that are sold with
variable annuity (VA) products in the U.S. We provide numerical examples in
all sections and then conclude the paper in Section \#6 with some direction
for future research.

\section{Valuation of the Income Annuity}

In this section we very briefly review the valuation of single premium
immediate (income) annuities, mainly in order to introduce notation and
terminology for the remainder of the paper and provide background for those
unfamiliar with mortality-contingent claims. We refer the interested reader to
any basic actuarial textbook, such as Promislow (2006) or Milevsky (2006), for
the assumptions we gloss over.

The value of a life annuity that pays \$1 per annum in continuous-time, is
denoted by $\text{ALDA}(\tau;\rho,x)\,$, where $x$ denotes the purchase age, $\rho$
denotes the (insurance company) valuation discount rate and $\tau$ is the
start date. The ALDA is an acronym for \textsl{Advanced Life Deferred Annuity.}
When the ALDA start date is immediate $(\tau=0)$ we have the more familiar concept of
a \textsl{Single Premium Immediate Annuity}, whose value is
$\text{SPIA}(\rho,x):=\text{ALDA}(0;\rho,x)$. 
Either way, the annuity valuation factor is equal to:
\begin{equation}
\text{ALDA}(\tau;\rho,x):=E\left[  \int_{\tau}^{T_{x}}e^{-\rho t}dt\right]  =E\left[
\int_{\tau}^{\infty}1_{\{T_{x}>t\}}e^{-\rho t}dt\right]  =\int_{\tau}^{\infty
}\left.  _{t}p_{x}\right.  e^{-\rho t}dt, \label{ALDA.eq1}%
\end{equation}
where $T_{x}$ denotes the future lifetime random variable conditional on the
current (purchase) age $x$ of the annuitant and $(_{t}p_{x})$ denotes the
conditional probability of survival to age $(x+t)$. In the above expression
$\tau$ is deterministic and denotes the deferral period before the insurance
company begins making lifetime payments to the annuitant. It is an actuarial
identity that:%
\begin{equation}
\text{ALDA}(\tau;\rho,x):=\text{SPIA}(\rho, x+\tau)\times\left.  _{\tau}p_{x}\right.  \times
e^{-\rho\tau}, \label{ALDA.identity}%
\end{equation}
which is the product of the age--$(x+\tau)$ SPIA factor multiplied by the
conditional probability of surviving to age $(x+\tau)$ multiplied by the
relevant discount factor $e^{-\rho\tau}$. In other words, the cost of a
deferred annuity can be written in terms of an (older) immediate annuity, the
survival probability and the discount rate. This actuarial identity will be
used later when $\tau$ itself is randomized.

Note that the expectation embedded within equation (\ref{ALDA.eq1}) is taken
with respect to the physical (real world) measure underlying the distribution
of $T_{x}$, which, due to the law of large numbers and the ability to
eliminate all idiosyncratic mortality risk is also equal to the risk-neutral
measure. While outside the scope of this paper which deals exclusively with
complete markets, in the event the realized force of mortality itself is
stochastic, it may in fact generate a mortality risk premium in which case the
physical (real world) and risk-neutral measure might not be the same. We leave
this for other research.

Under any continuous law of mortality specified by a deterministic function
$\lambda_{t}>0$, the expectation in equation (\ref{ALDA.eq1}), the annuity
factor, can be re-written as:%
\begin{equation}
\text{ALDA}(\tau;\rho,x)=\int_{\tau}^{\infty}\,e^{-\int_{0}^{t}\lambda_{q}%
\,dq}e^{-\rho t}\,dt=\int_{\tau}^{\infty}e^{-\int_{0}^{t}(\lambda_{q}+\rho
)dq}dt. \label{ALDA.eq2}%
\end{equation}
For most of the numerical examples within the paper we will assume that
$\lambda_{t}$ obeys the Gompertz-Makeham (GM) law of mortality. The canonical
GM force of mortality (see the paper by Carri\`{e}re (1994) or Frees,
Carri\`{e}re and Valdez (1996) for example), can be represented by:%
\begin{equation}
\lambda_{t}=\lambda+\frac{1}{b}e^{\left(  \frac{x+t-m}{b}\right)  },
\label{GM.eq1}%
\end{equation}
where $\lambda\geq0$ is a constant non-age dependent hazard rate, $b>0$
denotes a dispersion coefficient and $m>0$ denotes a modal value. Our notation
for $\lambda_{t}$ assumes four embedded parameters: the current age $x$,
$\lambda,m$ and $b$. Note that when $m\rightarrow\infty$, and $b>0$, the GM
collapses to a constant force of mortality $\lambda$, and the future lifetime
random variable is exponentially distributed. We will obtain some limiting
expressions in this case. For the more general and practical GM law, our RCLA
valuation expressions will be stated as solutions to a PDE.

As far as the basic ALDA factor is concerned, in the case of GM mortality,
one can actually obtain a closed-form expression for equation (\ref{ALDA.eq2}%
), which -- to our knowledge -- was first suggested by Mereu (1962). See Milevsky (2006) 
for a derivation that:%

\begin{equation}
\text{ALDA}(\tau;\rho,x,\lambda,m,b)=\frac{b\Gamma(-(\lambda+\rho)b,\exp
\{\frac{x-m+\tau}{b}\})}{\exp\left\{  (m-x)(\lambda+\rho)-\exp\left\{
\frac{x-m}{b}\right\}  \right\}  }, \label{GOMA.factor}%
\end{equation}
where all the input variables are now explictely listed in the arguments of the ALDA function, 
and $\Gamma(x,y)$ denotes the incomplete Gamma function. The annuity factor
itself is a decreasing function of age $x$, deferral period $\tau$, and the
valuation rate $\rho$. To see this, Figure \#3 plots the annuity factor in
equation (\ref{GOMA.factor}), for a continuum of ages from $x=40$ to $x=80$
assuming the valuation rates, $\rho=3\%,5\%$ and $7\%$ and $\tau=0$ deferral
period.%
\begin{figure}
\begin{center}
\includegraphics[scale=.5]{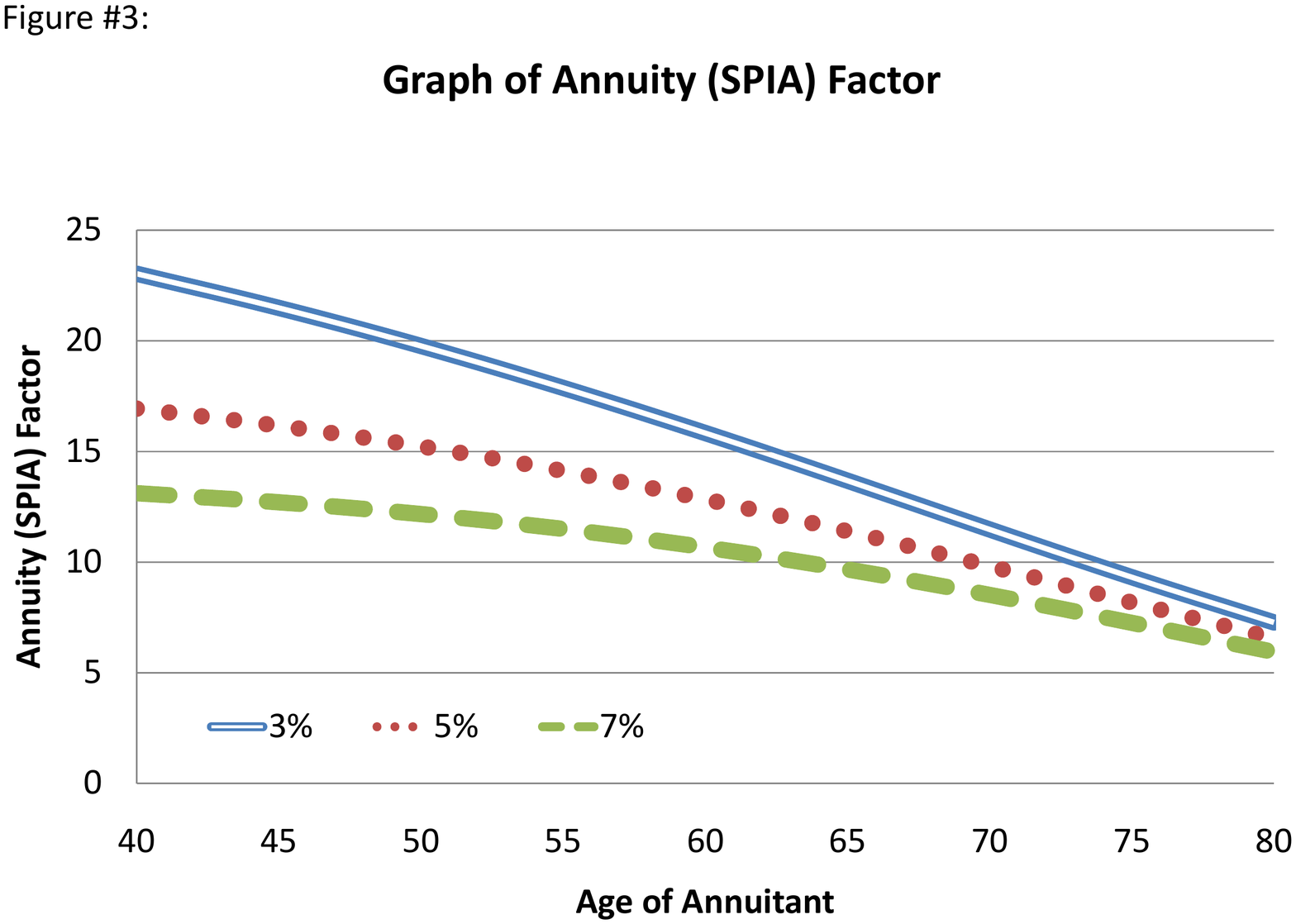} 
\label{figure3}
\end{center}
\end{figure}

Table \#1 displays some numerical values for a basic SPIA (immediate) and ALDA
(deferred) annuity factor, under the Gompertz Makeham $(m=86.3,b=9.5)$
continuous law of mortality. For example under an insurance valuation rate of
$\rho=5\%$, at the age of $x=40$, a buyer pays $\$16.9287$ for an income
stream of \$1 per year for life, starting immediately. If the annuity is
purchased at the same age but the start of income is delayed for $\tau=10$
years, the buyer pays $\$9.1010$ for \$1 per year for life, starting at age
50. In contrast, under the same $r=5\%$ rate, at age 65 the annuity value is
$\$11.3828$ per dollar of lifetime income, starting immediately and only
$\$4.0636$ if the start of the income is deferred for $\tau=10$ years. In
general, for higher valuation rates, advanced ages and longer deferral periods,
the annuity factor is lower. Note that the above Gompertz-Makeham assumptions
imply the conditional expectation of life at age 65 is 18.714 
years, which can be easily obtained by substituting an
insurance valuation rate of $\rho=0\%$ into the annuity factor. Note that no
death benefits or guarantee periods are assumed in these valuation
expressions. Thus, the occurrence of death prior to the end of the deferral
period will result in a complete loss of premium.
\begin{table}
\textbf{Table \#1}
\begin{center}%
\begin{tabular}
[c]{||c||c||c||c||c||}\hline\hline
\multicolumn{5}{||c||}{\textbf{Value of Single Premium Immediate/Deferred
Annuity}}\\\hline\hline
Purchase Age & Deferral & $\rho=3\%$ & $\rho=5\%$ & $\rho=7\%$\\\hline\hline
Age = 40 & $\tau=0$ yrs. & \$23.0144 & \$16.9287 & \$13.1126\\\hline\hline
& $\tau=5$ yrs. & \$18.3822 & \$12.5148 & \$8.9034\\\hline\hline
& $\tau=10$ yrs. & \$14.4228 & \$9.1010 & \$5.9575\\\hline\hline
& $\tau=20$ yrs. & \$8.2124 & \$4.4665 & \$2.4877\\\hline\hline
Age = 50 & $\tau=0$ yrs. & \$19.7483 & \$15.2205 & \$12.1693\\\hline\hline
& $\tau=5$ yrs. & \$15.1364 & \$10.8256 & \$7.9778\\\hline\hline
& $\tau=10$ yrs. & \$11.2448 & \$7.4697 & \$5.0815\\\hline\hline
& $\tau=20$ yrs. & \$5.3714 & \$3.0815 & \$1.7921\\\hline\hline
Age = 65 & $\tau=0$ yrs & \$13.6601 & \$11.3828 & \$9.6609\\\hline\hline
& $\tau=5$ yrs. & \$9.1653 & \$7.0974 & \$5.5719\\\hline\hline
& $\tau=10$ yrs. & \$5.6499 & \$4.0636 & \$2.9515\\\hline\hline
& $\tau=20$ yrs. & \$1.3886 & \$0.8577 & \$0.5320\\\hline\hline
Age = 75 & $\tau=0$ yrs. & \$9.2979 & \$8.1680 & \$7.2460\\\hline\hline
& $\tau=5$ yrs. & \$5.0620 & \$4.1250 & \$3.3839\\\hline\hline
& $\tau=10$ yrs. & \$2.2852 & \$1.7240 & \$1.3062\\\hline\hline
& $\tau=20$ yrs. & \$0.1645 & \$0.1055 & \$0.0677\\\hline\hline
\end{tabular}
\newline
\end{center}

Note: Table displays the basic annuity factor -- with no market contingencies
-- which is the actuarial present value per \$1 of annual income (in
continuous time) for life. The mortality is assumed Gompertz with parameters
$\lambda=0$, $m=86.3$ and $b=9.5$. Prices are risk neutral (ie. $\mu=\rho=r=$
risk-free rate). No death benefits or guarantee periods are assumed. Thus, a
death prior to the end of the deferral period will result in a complete loss of
premium. 
\end{table}

The ruin-contingent life annuity and its variants, which we will formally
define in the next section, can be viewed as generalizations of the ALDA
factor, but where the deferral period $\tau$ is {\it stochastic} and tied
to the performance of a reference portfolio index.

\section{Retirement Spending and Lifetime Ruin}

The RCLA is an income annuity that begins payment when a reference portfolio
index (RPI) hits zero, or is ruined. In this section we describe the mechanics
of the state variable which triggers the payment. To begin with we assume
investment returns are generated by a lognormal distribution so that the
RPI\ obeys the classic ``workhorse'' of financial economics:%
\begin{equation}
dW_{t}=\left(  \mu W_{t}-\gamma I_{0}\right)  dt+\sigma W_{t}dB_{t},\qquad
W_{0}=I_{0}. \label{RPI.eq1}%
\end{equation}
The parameter $\mu$ denotes the drift rate and $\sigma$ denotes the diffusion
coefficient. The constant $\gamma I_{0}$ denotes the annual spending rate
underlying the RPI. Note that when $\gamma=0$ the process $W_{t}$ collapses to
a geometric Brownian motion (GBM) which can never access zero in finite time.
The presence of $\gamma$ reduces the drift and makes zero accessible in finite
time. The greater the value of $\gamma$ the greater is the probability, all
else being equal, that $W_{t}$ hits zero\footnote{The evolution of retirement
wealth implied by equation (\ref{RPI.eq1}) is often studied as an alternative
to annuitization in the pension and retirement planning literature. See, for
example, the paper by Albrecht and Maurer (2002) or Kingston and Thorp (2005),
in which $\gamma I_{0}$ is set equal to the relevant SPIA factor times the
initial wealth at retirement.}.

We define the ruin time $R$ of the diffusion process as a hitting-time or
level-crossing time, which should be familiar from the classical insurance or
queueing theory literature. Formally it is defined as:%
\begin{equation}
R:=\inf\left\{  t;W_{t}\leq0\mid W_{0}=I_{0}\right\}  . \label{ruin.time}%
\end{equation}
There is obviously the possibility that $R=\infty$ and the RPI never hits
zero. See the paper by Huang, Milevsky and Wang (2004) or the paper by Dhaene,
Denuit, Goovaerts, Kaas and Vyncke (2002), as well as Norberg (1999), for a
detailed and extensive description of the various analytic and moment-matching
techniques that can be used to compute the probability distribution of $R$.
Likewise, see the paper by Young (2004) for a derivation of asset allocation
control policies on $(\mu,\sigma)$ that can be used to minimize ruin
probabilities within the context of retirement spending. Our focus is not on
controlling $R$ or explicitly estimating $\Pr[T_{x}\geq R]$ which is the
lifetime ruin probability. We are simply interested in using $R$ as a deferral
time for an income annuity.


\section{The Ruin-Contingent Life Annuity (RCLA)}

Like the generic annuity, the ruin-contingent life annuity (RCLA) is acquired
with a lump-sum premium now, and eventually pays \$1 of income per year for
life. However, the income payments do not commence until time $\tau=R$, when
the reference portfolio index (RPI) hits zero. And, if the RPI never hits zero
-- or the annuitant dies prior to the RPI hitting zero -- the RCLA expires
worthless. Thus, the defining structure of the RCLA is similar to the annuity
factor in equation (\ref{ALDA.eq1}), albeit with a stochastic upper \emph{and}
lower bound:%
\begin{equation}
\text{RCLA}(I_{0};\rho,x,\lambda,m,b,\gamma,\mu,\sigma,\tau)=E\left[
{\displaystyle\int_{R}^{R\vee T_{x}}}
e^{-\rho t}dt\right]  \label{RCLA.eq1}%
\end{equation}
The \$1 of annual lifetime income starts at time $R$ and continues until time
$\max\{R,T_{x}\}$. Thus, if the state-of-nature is such that $T_{x}<R$, and
the annuitant is dead prior to the ruin time, the integral from $R$ to $R$ is
zero and the payout is zero. Each RCLA unit entitles the annuitant to \$1 of
income. Thus, if one thinks of an RCLA as \textquotedblleft
insuring\textquotedblright\ a $\gamma I_{0}$ drawdown plan, then buying
$\gamma I_{0}$ RCLA units, would continue to pay $\gamma I_{0}$ dollars upon ruin.

Now, in order to derive a valuation relationship for the RCLA defined by
equation (\ref{RCLA.eq1}) we do the following. First, we simplify notation by
writing the annuity factor $\text{ALDA}(\xi;\rho,x)$ as $F(\xi)$. In other words,
\begin{equation}
F(\xi)=\int_{\xi}^{\infty}\left.  _{t}p_{x}\right.  e^{-\rho t}\,dt=E\left[
{\displaystyle\int_{\xi}^{\xi\vee T_{x}}}
e^{-\rho t}dt\mid\xi\right]  \label{RCLA.eq2}%
\end{equation}
Our problem then becomes to calculate:%
\begin{equation}
E[F(R)]=E\left[  E\Big[%
{\displaystyle\int_{R}^{R\vee T_{x}}}
e^{-\rho t}dt\mid R\Big]\right]  =E\left[  {\displaystyle\int_{R}^{R\vee
T_{x}}}e^{-\rho t}dt\right]  =\text{RCLA}(I_0). \label{RCLA.eq3}%
\end{equation}
Note that once again we rely on the law of large numbers -- from the
perspective of the insurance company -- to diversify away any idiosyncratic
longevity risk and value the RCLA based on (subjective, physical) mortality expectations.

Now, if $\mathcal{F}_{t}$ is the filtration generated by $W_{t}$, the
reference portfolio index, then $E[F(R)\mid \mathcal{F}_{t}]$ is a
martingale in $t$. By the Markov property, it can be represented in the form
$f(t\wedge R,W_{t})$, so applying Ito's lemma leads to the familiar
(Kolmogorov backward) PDE:%
\begin{equation}
f_{t}+(\mu w-\gamma I_{0})f_{w}+\frac{1}{2}\sigma^{2}w^{2}f_{ww}=0
\label{RCLA.eq4}%
\end{equation}
for $w>0$ and $t>0$. We now have an expression for (\ref{RCLA.eq1}) as
\begin{equation}
\text{RCLA}(I_0)=f(0,I_{0}). \label{initialcost}%
\end{equation}

Equation (\ref{RCLA.eq4}) differs from the famous Black-Scholes-Merton PDE by
the presence of the $\gamma I_{0}$ constant multiplying the space derivative
$f_{w}$. Also, our boundary conditions are different from the linear ones for
call and put options. Two of our boundary conditions are that
$f(t,w)\rightarrow0$ as either $t\rightarrow\infty$ or $w\rightarrow\infty$.
Intuitively, the RCLA is worthless in states of nature where the underlying
RPI never gets ruined, and/or only gets ruined after the annuitants have all
died. The boundary condition we require is that $f(t,0)=F(t)$, defined by
equation (\ref{RCLA.eq2}). The intuition here is that if-and-when the RPI hits
zero at some future time $\xi$, a live annuitant will be entitled to lifetime
income whose actuarially discounted value is the annuity factor $F(\xi)$.

Moreover, when $\lambda_{t}=\lambda$ is constant we recover the simple
expression $F(\xi)=e^{-(\lambda+\rho)\xi}/(\lambda+\rho)$ and one can simplify
the entire problem to obtain a solution of the form $f(t,w)=e^{-(\lambda
+\rho)t}h(w)$, where the new one-dimensional function $h(w)$ satisfies the
ODE:
\begin{equation}
(\mu w-\gamma I_{0})h_{w}(w)+\frac{1}{2}\sigma^{2}w^{2}h_{ww}(w)-(\lambda
+\rho)h(w)=0, \label{RCLA.ode}%
\end{equation}
where $h_{w}$ and $h_{ww}$ denote the first and second derivatives
respectively. The two boundary conditions are $h(\infty)=0$ and
$h(0)=1/(\lambda+\rho)$. But, when $\lambda_{t}$ is non-constant and obeys the
full GM law, we must use the full expression $F(\xi)=\left.  _{\xi}%
\overline{a}_{x}(\rho)\right.  $ for the boundary condition, which was
displayed in equation (\ref{GOMA.factor}). Note that we then have a parabolic
PDE, which can be solved numerically.

Note that in both equations (\ref{RCLA.eq4}) and (\ref{RCLA.ode}) we maintain
a distinction between the drift rate $\mu$ and the insurance valuation rate
$\rho$. One reason for doing so is to leave open the possibility of using our
valuation equation to calculate the expected RCLA returns under the physical
measure, in which $\mu$ could be the growth rate under the physical measure
even if $\rho=r$ is the risk-free interest rate. Another reason is that even
if we are interested in calculating prices (or the costs of manufacturing or
hedging the products), and so take $\mu=r$ to be the risk-free interest rate,
an RCLA contract could still in principal specify a different value for the
insurance valuation rate $\rho$. We will discuss this further in Section
\ref{hedgingsection}. However, in our numerical examples below we will take
$\mu=\rho=r$ (the risk-free rate) as in the Black-Scholes-Merton economy, etc.

There are also extensions of this analysis that should be possible. It would
be natural, given this product's role in retirement savings, to incorporate
real inflation adjustment factors into the RCLA payouts. Since the product is
envisioned as having a long horizon, it would also be worthwhile to
incorporate stochastic volatility into the model for the underlying asset
price, as well as variable interest rates. Finally, we have assumed complete
diversification of mortality risk, due to the law of large numbers and the
sale of a very large number of contracts. This is only a first approximation
to actuarial practice, in which adjustments are made to account for the
non-zero mortality risk still present when only a finite number of contracts
are sold. We hope to treat several of these effects in subsequent work, but
note that in some cases this means moving to techniques suitable for
incomplete markets.

\subsection{Solution Technique}

To solve the PDE for $f(t,w)$ which is displayed in equation (\ref{RCLA.eq4}),
we first use the following transformation:
\begin{equation}
f(t,w)=F^{\prime}(t)u(t,w), \label{solution.eq1}%
\end{equation}
where without any loss of generality $u(t,w)$ is defined as a new (possibly)
two-dimensional function. By taking partial derivatives and the chain rule, it
is easy to verify that:
\begin{equation}
f_{t}=F^{\prime\prime}u+F^{\prime}u_{t},\;f_{w}=F^{\prime}u_{w},\;f_{ww}%
=F^{\prime}u_{ww}, \label{solution.eq2}%
\end{equation}
where once again we use shorthand notation $f_{t},f_{w}$ and $f_{ww}$ for the
three derivatives of interest. By substituting equation (\ref{solution.eq2})
into equation (\ref{RCLA.eq4}), the valuation PDE for $f(t,w)$ can be written
in terms of the known function $F(t)$ and the yet-to-be-determined function
$u(t,w)$ as:
\begin{equation}
\frac{F^{\prime\prime}}{F^{\prime}}u+(\mu w-\gamma I_{0})u_{w}+\frac{1}%
{2}\sigma^{2}w^{2}u_{ww}+u_{t}=0. \label{solution.eq3}%
\end{equation}
Now, since by construction,
\begin{equation}
F(\xi)=\int_{\xi}^{\infty}e^{-\int_{0}^{s}(\lambda_{q}+\rho)dq}ds,
\label{solution.eq4}%
\end{equation}
we have that
\begin{equation}
F^{\prime}(\xi)=-e^{-\int_{0}^{\xi}(\lambda_{q}+\rho)dq}ds,\;F^{\prime\prime
}(\xi)=-(\lambda_{\xi}+r)F^{\prime}(\xi). \label{solution.eq5}%
\end{equation}
Thus, expressed in units of time $t$, the PDE for $u(t,w)$ becomes
\begin{equation}
-(\lambda_{t}+\rho)u+(\mu w-\gamma I_{0})u_{w}+\frac{1}{2}\sigma^{2}%
w^{2}u_{ww}+u_{t}=0, \label{solution.eq6}%
\end{equation}
where $u$ is shorthand for $u(t,w)$, and $u_{t},u_{w},u_{ww}$ are shorthand
notations for the time, space and second space derivatives, respectively. Now,
going back to the decomposition of $f(t,w)$ in equation (\ref{solution.eq1}),
and using the boundary condition for $f(t,w)$ at $w=0$, we have
\begin{equation}
F(t)=f(t,0)=F^{\prime}(t)u(t,0), \label{solution.eq7}%
\end{equation}
and
\begin{equation}
F^{\prime}(t)=F^{\prime}(t)u_{t}(t,0)+F^{\prime\prime}(t)u(t,0),
\label{solution.eq8}%
\end{equation}
from which we obtain
\begin{equation}
u_{t}(t,0)=(\lambda_{t}+\rho)u(t,0)+1. \label{solution.eq9}%
\end{equation}
For the numerical procedure, we first generate values of $u(t,w)$ by solving
equation (\ref{solution.eq6}) with boundary conditions from equation
(\ref{solution.eq9}) and condition $u(w,t)\rightarrow0$ as $w\rightarrow
\infty$ and $t\rightarrow\infty$. Then we multiply $u(t,w)$ by $F^{\prime}(t)$
to generate the RCLA values of $f(t,w)$.
\addtocounter{figure}{3}
\begin{figure}
\begin{center}
\includegraphics[scale=.5]{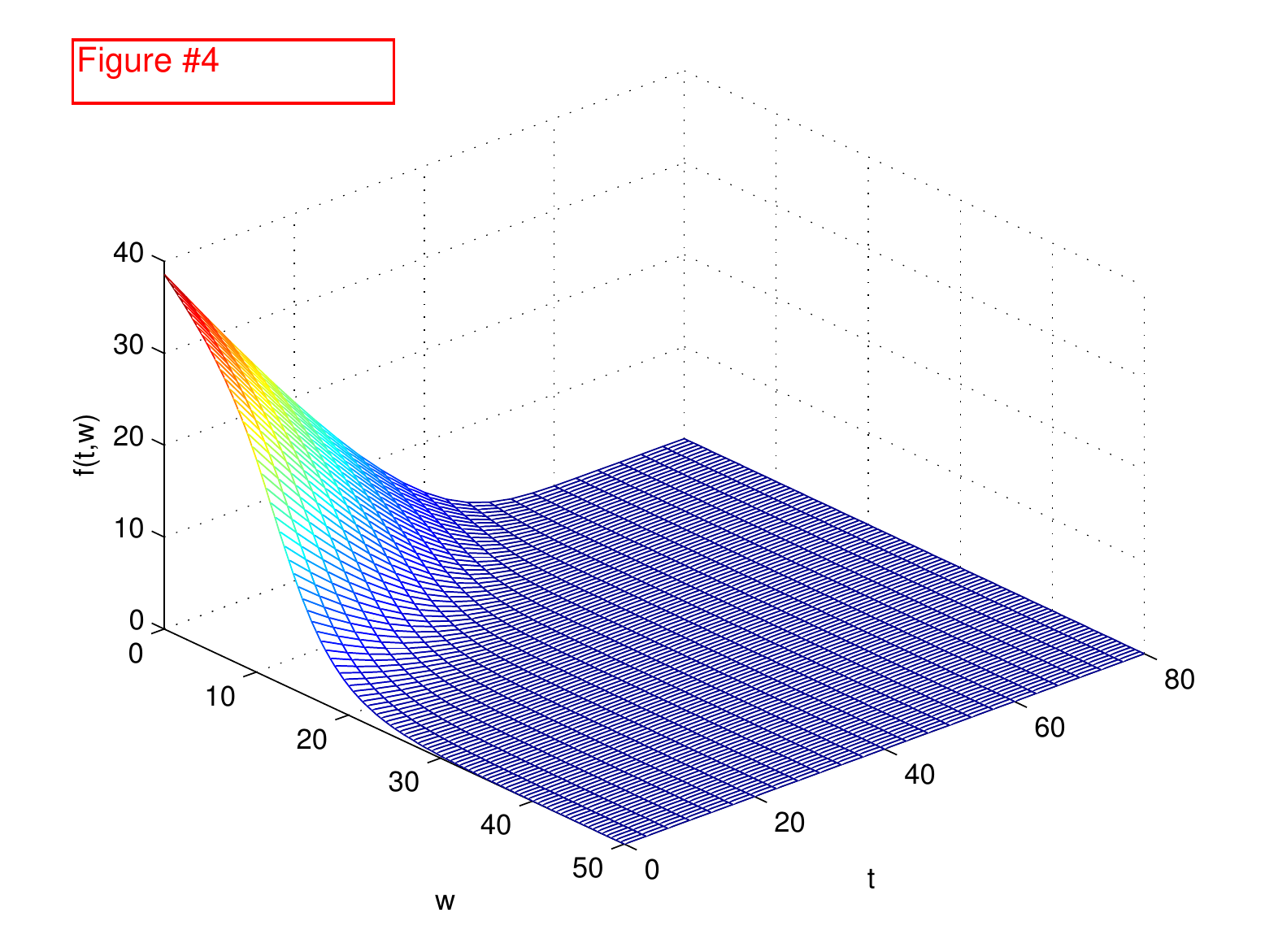} 
\caption{RCLA values}
\label{figure4}
\end{center}
\end{figure}

If necessary, values can also be calculated simultaneously for multiple values
of $\gamma$ by rescaling. This is the case, for example, in the numerical
examples and tables found below. We let $\tilde{w}=w/\gamma I_{0}$ and define
$\tilde{u}(t,\tilde{w})=u(t,w)$. Then the PDE for $\tilde{u}$ is seen to be
\begin{equation}
-(\lambda_{t}+\rho)\tilde{u}+(\mu\tilde{w}-1)\tilde{u}_{\tilde{w}}+\frac{1}%
{2}\sigma^{2}\tilde{w}^{2}\tilde{u}_{\tilde{w}\tilde{w}}+\tilde{u}_{t}=0,
\label{solution.eq10}%
\end{equation}
with the same boundary conditions as before. The parameter $\gamma$ no longer
appears, so only one PDE needs to be solved, after which we can calculate
\begin{equation}
f(t,w)=F^{\prime}(t)u(t,w)=F^{\prime}(t)\tilde{u}\Big(t,\frac{w}{\gamma I_{0}%
}\Big)
\end{equation}
for any desired value of $\gamma$. In fact, we will drop the \textquotedblleft
tilde\textquotedblright\ notation, since $\tilde{u}$ is just $u$ in the
special case $\gamma I_{0}=1$. Thus, if we have computed that particular
function $u$ we then get RCLA values for other $\gamma$'s as
\begin{equation}
\text{RCLA}(I_0)=F^{\prime}(0)u(0,I_{0}/\gamma I_{0})=F^{\prime}(0)u(0,1/\gamma).
\end{equation}

In Figure \#4 we plot $f(t,w)$, which is the RCLA value, assuming $\mu
=\rho=r=0.06$ (i.e. for risk neutral pricing) $m=86.3$, $b=9.5$ and $x=40$
(all three embedded mortality parameters) and $\lambda=0.003$, which is
the age-independent component of the Gompetz-Makeham law, and finally $\sigma=0.1$. 
The computation is done by solving the equation for $u(t,w)$ for $0<t<80$ (corresponding to a
maximum age of death of 120) and $0<w<50$, and using a normalized value of
$\gamma I_{0}=1$. As mentioned above, the function $f(t,w)$ is recovered by
multiplying $u(t,w)$ by $F^{\prime}(t)$, evaluated by numerical quadrature
based on Simpson's rule. We can then use $f$ to value RCLA's with different
withdrawal rates. Thus, for example, the point $f(0,10)$ corresponds to the
price of a \$1 per year for life RCLA, purchased at the age of $40$, assuming
a spending rate of $\gamma=1/10=10\%$ of the RPI level $I_{0}=100$.

Note that we experimented with different domain sizes up to $w=100$ and no
visible differences in results were observed, relative to the case when
$w=50$. (A single-run took a few seconds for a grid resolution of $\delta
w=0.1$ on a MacBook Pro.) Note that when $\lambda_{t}=\lambda$ is a constant,
$u_{t}=0$ and we recover the above-referenced special case mentioned prior to
equation (\ref{RCLA.ode}).

\subsection{Numerical Examples}

Table \#2a\textbf{ }displays the (risk neutral) value of the RCLA -- which
pays \$1 per year of lifetime income -- assuming the Reference Portfolio Index
(RPI) is allocated to LOW volatility investments with $\sigma=10\%$. The
spending $\gamma$ denotes the fixed percentage of the initial RPI level
$I_{0}$ that is withdrawn annually (and in continuous time) until ruin. When
$\gamma=\infty$ the RPI\ hits zero immediately and the RCLA collapses to a
basic annuity priced in Table \#1. The mortality is assumed Gompertz with
parameters $m=86.3$ and $b=9.5$. Thus, for example, at the age of 65 the value
of a 5\% withdrawal RCLA on a \textquotedblleft low
volatility\textquotedblright\ index is $\$0.6872$ under a valuation rate of
$\rho=3\%$ and a mere $\$0.1384$ under a valuation rate of $\rho=5\%.$ In
fact, even at the young age of $x=50$ and under a relatively high spending
percentage of $\gamma=7\%$, the value of the RCLA is only $\$2.4921$ per
dollar of lifetime income upon ruin, under the $5\%$ valuation rate.
Predictably, at advanced ages the same 7\% withdrawal RCLA is valued at only a
fraction of this cost. For example, at age $x=75,$ and under a valuation rate
$\rho=5\%,$ the value of the RCLA is only\ $\$0.1965$. This is the impact of
low ($\sigma=10\%$) investment volatility; naturally when $\sigma$ and
$\gamma$ are low, the probability of lifetime ruin is very small. In contrast,
Table \#2b\textbf{ }which is identical in structure to \#2a displays the (risk
neutral) value of the RCLA assuming the Reference Portfolio Index (RPI) is
allocated to high volatility investments with $\sigma=25\%$. Once again the
RPI spending rate $\gamma$ denotes the fixed percentage withdrawn.
\begin{table}
\textbf{Table \#2a}
\begin{center}%
\begin{tabular}
[c]{||c||c||c||c||c||}\hline\hline
\multicolumn{5}{||c||}{\textbf{Ruin-Contingent Life Annuity (RCLA): LOW
Volatility (}$\sigma=10\%$)}\\\hline\hline
\multicolumn{5}{||c||}{Reference Portfolio Index (RPI) Initial Value is
$W_{0}=I_{0}=\$100$}\\\hline\hline
\multicolumn{5}{||c||}{Lifetime Payout Upon Ruin is \$1 per year}%
\\\hline\hline
Initial Purchase & RPI Spending & $\rho=3.0\%$ & $\rho=5.0\%$ & $\rho
=7.0\%$\\\hline\hline
Age = 50 & $\gamma=\infty$ & \$19.7483 & \$15.2205 & \$12.1693\\\hline\hline
& $\gamma=10\%$ & \$10.0297 & \$5.5770 & \$2.7307\\\hline\hline
& $\gamma=7\%$ & \$6.3444 & \$2.4921 & \$0.6928\\\hline\hline
& $\gamma=6\%$ & \$4.6797 & \$1.4549 & \$0.2887\\\hline\hline
& $\gamma=5\%$ & \$2.9226 & \$0.6470 & \$0.0820\\\hline\hline
& $\gamma=4\%$ & \$1.3642 & \$0.1853 & \$0.0129\\\hline\hline
& $\gamma=3\%$ & \$0.3716 & \$0.0249 & \$0.0008\\\hline\hline
Age = 65 & $\gamma=\infty$ & \$13.6601 & \$11.3828 & \$9.6609\\\hline\hline
& $\gamma=10\%$ & \$4.7321 & \$2.6623 & \$1.2869\\\hline\hline
& $\gamma=7\%$ & \$2.2498 & \$0.8381 & \$0.2217\\\hline\hline
& $\gamma=6\%$ & \$1.3972 & \$0.4024 & \$0.0758\\\hline\hline
& $\gamma=5\%$ & \$0.6872 & \$0.1384 & \$0.0168\\\hline\hline
& $\gamma=4\%$ & \$0.2294 & \$0.0282 & \$0.0019\\\hline\hline
& $\gamma=3\%$ & \$0.0385 & \$0.0024 & \$0.0001\\\hline\hline
Age = 75 & $\gamma=\infty$ & \$9.2979 & \$8.1680 & \$7.2460\\\hline\hline
& $\gamma=10\%$ & \$1.7928 & \$0.9691 & \$0.4433\\\hline\hline
& $\gamma=7\%$ & \$0.5818 & \$0.1965 & \$0.0476\\\hline\hline
& $\gamma=6\%$ & \$0.2930 & \$0.0752 & \$0.0130\\\hline\hline
& $\gamma=5\%$ & \$0.1094 & \$0.0194 & \$0.0022\\\hline\hline
& $\gamma=4\%$ & \$0.0253 & \$0.0027 & \$0.0002\\\hline\hline
& $\gamma=3\%$ & \$0.0026 & \$0.0001 & \$0.0000\\\hline\hline
\end{tabular}
\newline\newline
\end{center}

Notes: Table displays the value of the RCLA -- which pays \$1 per year of
lifetime income -- assuming the Reference Portfolio Index (RPI) is allocated
to LOW volatility investments with $\sigma=10\%$. The spending $\gamma$
denotes the fixed percentage of the initial RPI level $I_{0}$ that is
withdrawn annually (and in continuous time) until ruin. When $\gamma=\infty$
the RPI\ hits zero immediately and the RCLA collapses to a basic annuity
displayed in Table \#1. The mortality is assumed Gompertz with parameters
$\lambda=0$, $m=86.3$ and $b=9.5$. Prices are risk neutral (ie. $\mu=\rho=r=$
risk-free rate).
\end{table}

\begin{table}
\textbf{Table \#2b}

\begin{center}%
\begin{tabular}
[c]{||c||c||c||c||c||}\hline\hline
\multicolumn{5}{||c||}{\textbf{Ruin-Contingent Life Annuity (RCLA): High
Volatility (}$\sigma=25\%$)}\\\hline\hline
\multicolumn{5}{||c||}{Reference Portfolio Index (RPI) Initial Value is
$W_{0}=I_{0}=\$100$}\\\hline\hline
\multicolumn{5}{||c||}{Lifetime Payout Upon Ruin is \$1 per year}%
\\\hline\hline
Initial Purchase & RPI Spending & $\rho=3.0\%$ & $\rho=5.0\%$ & $\rho
=7.0\%$\\\hline\hline
Age = 50 & $\gamma=\infty$ & \$19.7483 & \$15.2205 & \$12.1693\\\hline\hline
& $\gamma=10\%$ & \$10.6454 & \$6.4788 & \$3.8827\\\hline\hline
& $\gamma=7\%$ & \$8.0694 & \$4.4234 & \$2.3422\\\hline\hline
& $\gamma=6\%$ & \$6.9858 & \$3.6383 & \$1.8159\\\hline\hline
& $\gamma=5\%$ & \$5.7793 & \$2.8227 & \$1.3093\\\hline\hline
& $\gamma=4\%$ & \$4.4570 & \$2.0038 & \$0.8466\\\hline\hline
& $\gamma=3\%$ & \$3.0457 & \$1.2249 & \$0.4571\\\hline\hline
Age = 65 & $\gamma=\infty$ & \$13.6601 & \$11.3828 & \$9.6609\\\hline\hline
& $\gamma=10\%$ & \$5.4652 & \$3.5491 & \$2.2451\\\hline\hline
& $\gamma=7\%$ & \$3.6732 & \$2.1443 & \$1.2009\\\hline\hline
& $\gamma=6\%$ & \$2.9976 & \$1.6622 & \$0.8790\\\hline\hline
& $\gamma=5\%$ & \$2.3015 & \$1.1972 & \$0.5899\\\hline\hline
& $\gamma=4\%$ & \$1.6103 & \$0.7719 & \$0.3477\\\hline\hline
& $\gamma=3\%$ & \$0.9645 & \$0.4144 & \$0.1657\\\hline\hline
Age = 75 & $\gamma=\infty$ & \$9.2979 & \$8.1680 & \$7.2460\\\hline\hline
& $\gamma=10\%$ & \$2.4354 & \$1.6324 & \$1.0625\\\hline\hline
& $\gamma=7\%$ & \$1.4095 & \$0.8470 & \$0.4882\\\hline\hline
& $\gamma=6\%$ & \$1.0713 & \$0.6113 & \$0.3330\\\hline\hline
& $\gamma=5\%$ & \$0.7531 & \$0.4031 & \$0.2049\\\hline\hline
& $\gamma=4\%$ & \$0.4705 & \$0.2322 & \$0.1082\\\hline\hline
& $\gamma=3\%$ & \$0.2420 & \$0.1072 & \$0.0445\\\hline\hline
\end{tabular}
\end{center}

Notes: Table -- which is identical in structure to \#2a -- displays the (risk
neutral) value of the RCLA assuming the Reference Portfolio Index (RPI) is
allocated to high volatility investments with $\sigma=25\%$. The RPI spending
rate $\gamma$ denotes the fixed percentage withdrawn.
\end{table}

Note the impact of the higher volatility rate on the RCLA value. The 5\%
withdrawal RCLA that cost $\$0.6872$ at the age of 65, under a valuation rate
of $\rho=3\%$ and low investment volatility in Table \#2a is now valued at
$\$2.3015$ in Table \#2b under an investment volatility of $\sigma=25\%$.
Similarly, the value for a 7\% withdrawal RCLA at age $x=7$ and under
$\rho=5\%$ quadruples to $\$0.8470.$

As one might expect intuitively, the value of an RCLA is also extremely
sensitive to the withdrawal percentage $\gamma$ underlying the RPI. For
example, at the age of 65 and under a valuation rate of $\rho=3\%$, a
withdrawal percentage of $\gamma=7\%$ on a high volatility RPI leads to an
RCLA value of $\$3.6732$, but is worth less than half at $\$1.6103$ under a
$\gamma=4\%$ withdrawal percentage. One can interpret these results as
indicating that insuring lifetime income against ruin at a 7\% withdrawal rate
is roughly $125\%$ more expensive than insuring against ruin at a 4\%
withdrawal rate. This provides an economic benchmark by which different
spending strategies can be compared.

\subsection{Hedging \label{hedgingsection}}

Our price, determined by risk-neutral valuation in previous sections,
represents a hedging cost. It is worth making the hedging argument explicit
(and evaluating Delta), even though this has certainly been implicit in what
we described above.

The partial differential equations given in the preceding sections evaluate
expectations. In the complete markets setting, the expectations are
risk-neutral, and represent hedging costs. In that setting, we normally choose
the equity growth rate $\mu$ and the insurance valuation rate $\rho$ to both
coincide with the risk free interest rate $r$: $\mu=\rho=r$. This is the
setting used for the numerical examples given above. But we could also use the
PDE's to work out discounted expected cash flows under the real-world or
physical measure, a problem that can arise in aspects of risk management other
than pricing. In that case we would apply the above formulas with $\mu$
equalling the real-world equity growth rate, and $\rho=r$ to be the risk-free rate.

By generalizing the RCLA slightly, we can also imagine using the PDE when
$\mu=r$ (so our measure is risk-neutral and we're looking at pricing and
hedging), but $\rho<r$. As we shall see below, this would be the case if
payments from the RCLA were not fixed at \$1 per year for life, but rather at
$e^{\delta t}$, where $\delta=\mu-\rho$. This would correspond to an
inflation-enhanced RCLA in which a fixed inflation rate $\delta$ is
incorporated into the contract, so payments increase over time at rate
$\delta$. The standard RCLA described earlier is just the case $\delta=0$. In
this subsection (and this subsection only) we will work out the hedging
portfolio assuming a complete market, with risk-free rate $\mu=r$ and a
valuation rate $\rho=r-\delta$. We do not change the definition of the
reference portfolio.

Note that we do not hedge the RCLA ``derivative'' using the reference
portfolio index (RPI) $W_{t}$, satisfying $dW_{t}=(rW_{t}-\gamma
I_{0})\,dt+\sigma W_{t}\,dB_{t}$ and $W_{0}=I_{0}$, since that quantity
incorporates withdrawals and is not readily tradeable. Instead we use a stock
index $S_{t}$ without withdrawals (which is assumed tradeable), on which the
RPI is based. In other words,
\begin{equation}
dS_{t}=rS_{t}\,dt+\sigma S_{t}\,dB_{t}.
\end{equation}

We assume that a large number $N$ of RCLA's is sold at time 0, to age-$x$
individuals. The company hedges these with a portfolio worth $V_{t}$ at time
$t$. Then
\begin{equation}
V_{t}=\Delta_{t}S_{t}+\Psi_{t}%
\end{equation}
where $\Delta_{t}$ is the number of stock index units held, and $\Psi_{t}$ is a
position in a money market account with interest rate $r$. Since the number of
contracts is large, a predictable fraction ${}_{t}p_{x}$ of contract holders
are still alive at time $t$, leading to outflows from the hedging portfolio of
$e^{\delta_{t}}{}_{t}p_{x}N$, if ruin has occurred by time $t$. Thus
\begin{align}
dV_{t}  &  =\Delta_{t}\,dS_{t}+r\Psi_{t}\,dt-e^{\delta t}{}_{t}p_{x}%
N1_{\{R<t\}}\,dt\\
&  =rV_{t}\,dt+\Delta_{t}\sigma S_{t}\,dB_{t}-e^{\delta t}{}_{t}%
p_{x}N1_{\{R<t\}}\,dt.\nonumber
\end{align}
We obtain a positive solution by taking
\begin{equation}
V_{t}=Ne^{rt}f(t,W_{t})
\end{equation}
and
\begin{equation}
\Delta_{t}=%
\begin{cases}
Ne^{rt}W_{t}f_{w}(t,W_{t})/S_{t}, & W_{t}>0\\
0, & W_{t}=0.
\end{cases}
\end{equation}
The verification is a simple consequence of Ito's lemma, the fact that
$f(t,w)$ solves (\ref{RCLA.eq4}) when $w>0$, and the observation that
$Ne^{rt}f_{t}(t,0)=Ne^{rt}F^{\prime}(t)=-Ne^{rt}e^{-\rho t}{}_{t}%
p_{x}=-e^{\delta t}{}_{t}p_{x}N$. Put another way, the value of the stock
position in the hedge, per initial contract sold, is just
\begin{equation}
\Delta_{t}S_{t}/N=e^{rt}W_{t}f_{w}(t,W_{t}).
\end{equation}
This expression reflects the fact that our solution is written using $W_{t}$
rather than $S_{t}$, and the observation that $f$ is already a discounted
quantity (being a martingale). Note that the relation between $W_{t}$ and
$S_{t}$ could be made explicit, but is path dependent.

Finally, the initial hedging cost, per contract, is just $V_{0}/N=f(0,I_{0})$
as in (\ref{initialcost}). Of course, in reality a company would
simultaneously hedge a book of RCLA's with different purchase dates, and sold
to clients with a range of ages. But the above analysis serves to illustrate
the connection between hedging and pricing.

\section{More Exotic Time-Dependent Payouts}

We now describe two additional types of RCLA, both of which are motivated by
real-world products. In the first modification the spending rate $\gamma
I_{0}$ increases to $\gamma\max_{0\leq s\leq t}\{W_{s}\}$, which accounts for
good performance, each time the underling RPI reaches a new maximum. In other
words, this product could be used to insure a drawdown plan, in which
withdrawals ratchet or step up. At ruin time $R$, this product pays \$1 per
year for life akin to the generic RCLA. In the second modification the
spending rate increases in a similar manner, but the lifetime income -- which
starts upon the RPI's ruin time -- will be increased as well. Both of these
RCLA variants are embedded within the latest generation of variable annuity
(VA) policies sold around the world with guaranteed lifetime withdrawal
benefits (GLWB). We now proceed to describe and value them in detail.

\subsection{The Fast-RCLA}

Once again, we let $T_{x}$ denote the remaining lifetime random variable under
a deterministic hazard rate $\lambda_{t}$, and we assume the RPI process
$W_{t}$ is independent of $T_{x}$ and satisfies the following diffusion
equation:%
\begin{equation}
dW_{t}=(\mu W_{t}-g(t,M_{t}))dt+\sigma W_{t}dB_{t},\qquad W_{0}=I_{0}
\label{General.RPI}%
\end{equation}
where the new function $M_{t}$ is defined as:%
\begin{equation}
M_{t}=\max_{0\leq s\leq t}W_{s}. \label{FRCLA.eq1}%
\end{equation}
Both $W_{t}$ and $M_{t}$ are now defined up until the time $R$ that $W_{t}$
hits zero. Note that the drift term in equation (\ref{General.RPI}) now
includes a more general specification and is not necessarily a constant
deterministic term $\gamma I_{0},$ as in the basic RCLA case. The modified
product that we call a \emph{Fast} RCLA differs from the basic RCLA in that
the spending\ function is defined in the following manner.%

\begin{equation}
g(t,m)=\left\{
\begin{array}
[c]{cc}%
0 & t\leq\tau\\
\gamma\max\{m,W_{0}e^{\beta\tau}\} & t>\tau
\end{array}
\right.  , \label{FRCLA.eq2}%
\end{equation}
where the new constant $\beta$ denotes a ``bonus rate'' for delaying $\tau$
years prior to spending/withdrawals. Note that $\tau$ is now a deferral period
before the RPI begins withdrawals. The constant $\gamma$ multiplying the
$\max$ function in equation (\ref{FRCLA.eq2}) serves the same role as $\gamma$
in the basic RCLA. It is a pre-specified percentage rate of some initial RPI value.

Thus, for example, assume that $W_{0}=I_{0}=100$ and that during the first ten
years $(t\leq\tau=10)$ the reference portfolio index $W_{t}$ grows at some
(lognormally distributed) rate and without any withdrawals. Then, after ten
years $(t>\tau=10)$ the RPI starts to pay-out the greater of (i) $\gamma=5\%$
of the the maximum RPI value $M_{10}$ observed to date, and (ii) $\gamma=5\%$
of $100e^{(0.05)10}=\allowbreak164.\,\allowbreak87$, which is \$8.2 per year.
Then, each time the process $W_{t}$ reaches a new high, so that $M_{t}=W_{t}$,
the spending rate $g(t,M_{t})$ is reset to $(0.05)W_{t}=(0.5)M_{t}$. Then,
if-and-when the RPI hits zero the insurance company makes payments of \$1 per
year for life, to the annuitant.

The value of the Fast RCLA is (still) defined as:
\begin{equation}
\text{F-RCLA}(I_0;\rho,x,\lambda,m,b,I_{0},\gamma,\mu,\sigma,\tau,\beta
):=f(0,I_{0},I_{0}) \label{FRCLA.eq3}%
\end{equation}
where for $0<w\leq m$,
\begin{equation}
f(t,w,m)=E\left[  \int_{R}^{R\vee T_{x}}e^{-\rho s}ds\mid W_{t}=w,M_{t}%
=m\right]  .
\end{equation}
The only difference between the F-RCLA and the RCLA is in the structure of the
ruin time $R.$ When $\tau=0\,\ $and the RPI begins immediate withdrawals, the
(generic) F-RCLA is more expensive compared to a basic RCLA because the
ruin-time $R$ under the diffusion specified by equation (\ref{General.RPI})
will occur prior to (or at the same time) as the ruin-time generated by the
constant withdrawal implicit within equation (\ref{RPI.eq1}).

To solve this valuation equation we go back to the PDE for the basic RCLA
which we derived in the previous section. Note that the original PDE,
displayed in equation (\ref{RCLA.eq4}), did not involve the hazard rate
function $\lambda_{t}$. Rather, the mortality was embedded into the boundary
conditions. We take advantage of the same idea for the Fast-RCLA.

First, we tinker with the definition of the $g(t,m)$ spending function. We
re-scale by starting $M_{t}$ at $W_{0}e^{\beta\tau}$ rather than at $W_{0}$.
So let $\overline{M}_{t}=W_{0}e^{\beta\tau}\vee\max_{0\leq s\leq t}W_{s}$. We
then define a \textquotedblleft moneyness\textquotedblright\ variable
$Y_{t}=W_{t}/\overline{M}_{t}$, satisfying $0\leq Y_{t}\leq1$. Let $\bar
{g}(t,\overline{m})$ be $g(t,m)$ in terms of the new variables, so that:
\begin{equation}
\overline{g}(t,\overline{m})=%
\begin{cases}
0, & t\leq\tau\\
\gamma\overline{m}, & t>\tau.
\end{cases}
\label{FRCLA.eq4}%
\end{equation}

Our problem is now to calculate the value of a new function defined as%
\begin{equation}
h(t,y,\overline{m})=E[F(R)\mid Y_{t}=y,\overline{M}_{t}=m] \label{FRCLA.eq5}%
\end{equation}
where $F(\xi)$ is defined as above, and $R$ is the ruin time of $W_{t}$. Then
the F-RCLA value $f(t,w,m)=h(t,y,\overline{m})$ where $\overline{m}=m\vee
W_{0}e^{\beta\tau}$ and $w=y\overline{m}$.

The next step is to calculate $h$ using that%
\begin{equation}
E[F(R)\mid\mathcal{F}_{t}]=h(t\wedge R,Y_{t},\overline{M}_{t})
\label{FRCLA.eq6}%
\end{equation}
is a martingale. To apply Ito's lemma, we need to write down the stochastic
equations for $Y_{t}$ (the new moneyness variable) and $\overline{M}_{t}$ (the
new maximum diffusion value). Note that $\overline{M}_{t}$ is increasing, and
defining $dL_{t}=d\overline{M}_{t}/\overline{M}_{t}$, we have that $L_{t}$ is
a process that increases only when $Y_{t}=1$, and
\begin{equation}
d\overline{M}_{t}=\overline{M}_{t}\,dL_{t}. \label{FRCLA.eq7}%
\end{equation}
Likewise
\begin{align}
dY_{t}  &  =\frac{1}{\overline{M}_{t}}dW_{t}-\frac{W_{t}}{\overline{M}_{t}%
^{2}}d\bar{M}_{t}\label{FRCLA.eq8}\\
&  =\frac{\mu W_{t}-\overline{g}}{\overline{M}_{t}}\,dt+\frac{\sigma W_{t}%
}{\overline{M}_{t}}\,dB_{t}-\frac{W_{t}}{\overline{M}_{t}}\,dL_{t}\nonumber\\
&  =(\mu Y_{t}-\widehat{g}(t))\,dt+\sigma Y_{t}\,dB_{t}-Y_{t}\,dL_{t}\nonumber
\end{align}
where we use yet another function,%
\begin{equation}
\widehat{g}(t)=\frac{\overline{g}(t,\overline{m})}{\overline{m}}=%
\begin{cases}
0, & t\leq\tau\\
\gamma, & t>\tau.
\end{cases}
\label{FRCLA.eq9}%
\end{equation}
We interpret (\ref{FRCLA.eq8}) as a \textquotedblleft Skorokhod
equation\textquotedblright\ and $L_{t}$ as a \textquotedblleft local
time\textquotedblright\ of $Y_{t}$ at 1, the effect of which is to pull
$Y_{t}$ down when it reaches 1, to ensure that it does not ever exceed 1. In
particular, $L_{t}$ is determined by the process $Y_{t}$. Note that
$\overline{M}_{t}$ has now entirely disappeared from the stochastic equation
for $Y_{t}$, so in fact $Y_{t}$ is a one-dimensional Markov process all by
itself. Because $R$ is determined by $Y$, in fact%
\begin{equation}
h(t,y,\overline{m})=h(t,y) \label{FRCLA.eq10}%
\end{equation}
does not depend on $\overline{m}$ at all. We are able to make all of these
simplifications because of the simple structure of the original spending rate
$g(t,m)$ in equation (\ref{FRCLA.eq1}). If we had a more general withdrawal
rate, say of the form $g(t,w,m)$ where $g$ is a more complicated function than
the one used above, then we would have to keep track of the maximum state
variable $\overline{m}$ in addition to the moneyness state variable $y.$

Now, applying Ito's lemma, we get that for $t<R$,
\begin{equation}
dh(t,Y_{t})=[h_{t}+(\mu Y_{t}-\widehat{g}(t))h_{y}+\frac{1}{2}\sigma^{2}%
Y_{t}^{2}h_{yy}]\,dt+\sigma Y_{t}h_{y}\,dB_{t}-Y_{t}h_{y}\,dL_{t}.
\label{FRCLA.eq11}%
\end{equation}
For this to be a martingale, both the $dt$ and $dL_{t}$ terms must vanish. So
in particular,
\begin{equation}
h_{t}+(\mu y-\widehat{g}(t))h_{y}+\frac{1}{2}\sigma^{2}y^{2}h_{yy}=0
\label{FRCLA.eq12}%
\end{equation}
and $h_{y}=0$ when $y=1$ (recall that $dL_{t}=0$ unless $Y_{t}=1$). The latter
is one boundary condition, and $h(t,0)=F(t)$, $h(t,y)\rightarrow0$ as
$t\rightarrow\infty$ are the others. Note the similarity between the PDE we
must solve for the F-RCLA in equation (\ref{FRCLA.eq12}) and the original
valuation PDE for the RCLA displayed in equation (\ref{RCLA.eq4}). Besides the
boundary conditions, the only difference is that $\gamma I_{0}$ is replaced by
$\widehat{g}(t)$. So, in the Gompertz case there is one time variable and one
spatial variable.

\subsection{The Super-RCLA}

In the previously analyzed F-RCLA, the spending/withdrawal stepped-up over
time, but when ruin occurs the F-RCLA payout is the same as for the RCLA,
namely \$1 per year for life. This type of product is relevant in some
contexts but not in others. Sometimes the lifetime income that is promised
upon ruin can be greater than the originally guaranteed rate, and is linked to
the function $g(t,m)$ itself. Therefore, in this sub-section we examine the
case in which the lifetime income paid by the annuity is linked to the
increasing level of spending/withdrawals. As before, the RPI value satisfies
the process:
\begin{equation}
dW_{t}=(\mu W_{t}-g(t,M_{t}))\,dt+\sigma W_{t}\,dB_{t}, \label{SRCLA.eq1}%
\end{equation}
under the same $(\mu,\sigma)$ parameters and where the withdrawal function
$g(t,m)$ satisfies:
\begin{equation}
g(t,w,m)=%
\begin{cases}
0, & t<\tau\\
\gamma m, & t\geq\tau
\end{cases}
\label{SRCLA.eq2}%
\end{equation}
and $M_{t}=W_{0}e^{\beta\tau}\vee\max_{0\leq s\leq t}W_{s}$. Recall that
$\beta$ is a bonus rate (during the deferral period) and $\tau$ denotes the
length of deferral period, measured in years. In this sense, the underlying
diffusion and ruin-time dynamics are identical to the previously discussed
F-RCLA case.

However, in contrast to the \$1 of lifetime income payoff from the F-RCLA, we
define the {\em Super} RCLA value as:
\begin{align}
\text{S-RCLA}(I_0;\rho,x,\lambda,m,b,I_{0},\gamma,\mu,\sigma,\tau,\beta)  &
\text{:}=\frac{f(0,I_{0},I_{0})}{g(0,I_{0})}\label{SRCLA.eq3}\\
f(t,w,m)  &  :=E\left[  g(R,m)\int_{R}^{R\wedge T_{x}}e^{-\rho s}\,ds\right]
.\nonumber
\end{align}
The S-RCLA starts paying income for life when the process in equation
(\ref{SRCLA.eq1}) is ruined, but the income will not be \$1. Instead, it will
be equal to the withdrawal amount itself, $g(R,m)$, just prior to the time of
ruin $R$, divided by the initial withdrawal rate $g(0,I_{0})$. If there was no
step-up in the withdrawal spending prior to ruin, then the payout will simply
be \$1 for life, just like the F-RCLA and the original RCLA. We have decided
to define the function $f(t,w,m)$ so that we do not have to carry around the
denominator $g(0,I_{0})$ of equation (\ref{SRCLA.eq3}) during the entire derivation.

Either way, our boundary condition must change even though large parts of the
solution are similar to the F-RCLA and RCLA. We define the moneyness variable
$Y_{t}=W_{t}/M_{t}$ so that $0\leq Y_{t}\leq1$. Also, let $L_{t}$ be the local
time of $Y$ at 1, so
\begin{equation}
dY_{t}=(\mu Y_{t}-\widehat{g}(t))\,dt+\sigma Y_{t}\,dB_{t}-dL_{t}
\label{SRCLA.eq4}%
\end{equation}
where the (new) scaled variable $\widehat{g}(t)$ is now defined as:
\begin{equation}
\widehat{g}_{t}=%
\begin{cases}
0, & t<\tau\\
\gamma I_{0}, & t\geq\tau.
\end{cases}
\label{SRCLA.eq5}%
\end{equation}
By construction, we also have that $dM_{t}=M_{t}\,dL_{t}$. Moreover, the
S-RCLA value defined by:
\begin{equation}
E\left[  g(R,m)\int_{R}^{R\wedge T_{x}}e^{-\rho s}\,ds\mid\mathcal{F}%
_{t}\right]  \label{SRCLA.eq6}%
\end{equation}
will be a martingale. By the Markov property the S-RCLA value will be of the
form $f(t\wedge R,W_{t\wedge R},M_{t\wedge R})$ for some function $f$. There
is a scaling relationship $f(t,cw,cm)=cf(t,w,m)$, from which we conclude that
$f(t,w,m)=mh(t,y)$ for some function $h$ (where $y=w/m$). Applying Ito's
lemma,
\begin{equation}
d\left(  M_{t}h(t,Y_{t})\right)  =(h_{t}+h_{y}(\mu Y_{t}-\widehat{g}%
(t))+\frac{1}{2}\sigma^{2}Y_{t}^{2}h_{yy})\,dt+\sigma Y_{t}h_{y}\,dB_{t}%
+M_{t}(h-h_{y})\,dL_{t}. \label{SRCLA.eq7}%
\end{equation}
We conclude that
\begin{equation}
h_{t}+h_{y}(\mu y-\widehat{g}(t))+\frac{1}{2}\sigma^{2}y^{2}h_{yy}=0
\label{SRCLA.eq8}%
\end{equation}
for $0<y<1$, with boundary condition $h(t,1)=h_{y}(t,1)$ at $y=1$. There will
again be a boundary condition $h(t,w)\rightarrow0$ as $t\rightarrow\infty$. At
$y=0$ the boundary condition is that:
\begin{equation}
h(t,0)=%
\begin{cases}
0, & t<\tau\\
\gamma I_{0}F(t), & t\geq\tau
\end{cases}
\label{SRCLA.eq9}%
\end{equation}
where $F(t)$ is defined as before. Note that we are multiplying $\gamma I_{0}$
by the annuity factor $F(t)$ since the payoff is now specified in terms of the
spending rate and not single dollars. Also, since the equation is parabolic we
only need a boundary condition in time at $t=\infty$.

After solving this PDE for $h$, we recover the S-RCLA value as:%
\begin{equation}
f(0,w_{0},m_{0})=f(0,w_{0},w_{0}e^{\beta\tau})=w_{0}e^{\beta\tau}%
h(0,e^{-\beta\tau}).
\end{equation}
It is worth commenting on the boundary condition $h=0$ when $w=0$ and $t<\tau
$. This is because the formulation of the S-RCLA implies that the payout rate
$g(t,w,m)=0$ $\forall t$, if it happens that $R<\tau$. However, the RPI cannot
get ruined (in a GBM world) before time $\tau$: $P(R<\tau)=0$. So it is
presumably irrelevant what boundary condition we use when $w=0$ and $t<\tau$.%

\begin{table}
\textbf{Table \#3}

\begin{center}%
\begin{tabular}
[c]{||c||c||c||c||c||}\hline\hline
\multicolumn{5}{||c||}{\textbf{Super RCLA Value: Medium Volatility (}%
$\sigma=17\%$)}\\\hline\hline
\multicolumn{5}{||c||}{Lifetime Payout Upon Ruin is AT\ LEAST \$1 per
year}\\\hline\hline
Initial Purchase & Initial Spending Rate & $\rho=3.0\%$ & $\rho=5.0\%$ &
$\rho=7.0\%$\\\hline\hline
Age = 50 & $\gamma=10.0\%$ & \$13.1593 & \$8.4032 & \$5.2951\\\hline\hline
& $\gamma=7.0\%$ & \$10.6177 & \$6.0704 & \$3.2801\\\hline\hline
& $\gamma=5.5\%$ & \$8.5497 & \$4.3654 & \$2.0237\\\hline\hline
& $\gamma=4.0\%$ & \$5.6736 & \$2.3663 & \$0.8479\\\hline\hline
Age = 57 & $\gamma=10.0\%$ & \$10.2025 & \$6.6799 & \$4.2748\\\hline\hline
& $\gamma=7.0\%$ & \$7.7181 & \$4.4651 & \$2.4178\\\hline\hline
& $\gamma=5.5\%$ & \$5.8394 & \$2.9846 & \$1.3756\\\hline\hline
& $\gamma=4.0\%$ & \$3.4939 & \$1.4433 & \$0.5122\\\hline\hline
Age = 65 & $\gamma=10.0\%$ & \$6.6981 & \$4.4761 & \$2.8938\\\hline\hline
& $\gamma=7.0\%$ & \$4.5249 & \$2.6205 & \$1.4077\\\hline\hline
& $\gamma=5.5\%$ & \$3.0899 & \$1.5607 & \$0.7076\\\hline\hline
& $\gamma=4.0\%$ & \$1.5749 & \$0.6362 & \$0.2216\\\hline\hline
Age = 67 & $\gamma=10.0\%$ & \$5.8505 & \$3.9217 & \$2.5370\\\hline\hline
& $\gamma=7.0\%$ & \$3.8074 & \$2.1997 & \$1.1767\\\hline\hline
& $\gamma=5.5\%$ & \$2.5171 & \$1.2643 & \$0.5698\\\hline\hline
& $\gamma=4.0\%$ & \$1.2218 & \$0.4897 & \$0.1695\\\hline\hline
Age = 75 & $\gamma=10.0\%$ & \$2.8580 & \$1.9170 & \$1.2303\\\hline\hline
& $\gamma=7.0\%$ & \$1.5232 & \$0.8606 & \$0.4481\\\hline\hline
& $\gamma=5.5\%$ & \$0.8542 & \$0.4148 & \$0.1809\\\hline\hline
& $\gamma=4.0\%$ & \$0.3261 & \$0.1254 & \$0.0420\\\hline\hline
\end{tabular}

\end{center}

Notes: Table displays the value of the Super RCLA assuming the Reference
Portfolio Index (RPI) is allocated to medium volatility investments with
$\sigma=17\%$ volatility. The initial RPI spending $\gamma$ denotes the
percent of the initial index value that is withdrawn annually (in continuous
time). The factors in Table \#3 are not directly comparable to the factors in
Table \#2 since the lifetime income upon ruin could exceed \$1, if the RPI
``does well'' prior to ruin. The mortality is assumed Gompertz with parameters
$\lambda=0$, $m=86.3$ and $b=9.5$. Prices are risk neutral (ie. $\mu=\rho=r=$
risk-free rate).
\end{table}

Table \#3\textbf{ }displays the (risk neutral) value of the Super RCLA
assuming the Reference Portfolio Index (RPI) is allocated to medium volatility
($\sigma=17\%$) investments. The initial RPI spending $\gamma$ denotes the
percent of the initial index value that is withdrawn annually (in continuous
time.) The factors in Table \#3 are not directly comparable to the factors in
Table \#2 since the lifetime income upon ruin could exceed \$1, if the RPI
\textquotedblleft does well\textquotedblright\ prior to ruin. As an example,
consider a 67 year old with an initial spending rate of\ $\gamma=5.5\%$. Under
a valuation rate of $\rho=5\%$ and investment volatility of $\sigma=17\%$, an
S-RCLA guaranteeing a lifetime payout of at least \$1 upon ruin is valued at
$\$1.2643.$ Again, the actual guaranteed payout will be determined by the
extent of withdrawal step-ups during the spending period. In these examples,
Gompertz mortality is assumed with parameters $m=86.3$ and $b=9.5$.%

\begin{table}
\textbf{Table \#4}

\begin{center}%
\begin{tabular}
[c]{||c||c||c||c||c||}\hline\hline
\multicolumn{5}{||c||}{\textbf{Valuation of RCLA v.s. Super-RCLA under
Differing RPI Volatility (}$\rho=5\%$)}\\\hline\hline
\multicolumn{5}{||c||}{What is a Step-Up Really Worth?}\\\hline\hline
Purchase & Volatility ($\sigma$) & RCLA ($\gamma=5\%$) & SRCLA ($\gamma
=5\%$) & ``Super Premium''\\\hline\hline
Age = 57 & $\sigma=8\%$ & \$0.2102 & \$0.4341 & +\textit{106\%}\\\hline\hline
& $\sigma=15\%$ & \$0.8590 & \$1.9123 & +\textit{123\%}\\\hline\hline
& $\sigma=20\%$ & \$1.4378 & \$3.3569 & +\textit{133\%}\\\hline\hline
& $\sigma=25\%$ & \$2.0521 & \$5.0267 & +\textit{145\%}\\\hline\hline
Age = 62 & $\sigma=8\%$ & \$0.1096 & \$0.2233 & +\textit{104\%}\\\hline\hline
& $\sigma=15\%$ & \$0.5617 & \$1.2390 & +\textit{121\%}\\\hline\hline
& $\sigma=20\%$ & \$1.0088 & \$2.3325 & +\textit{131\%}\\\hline\hline
& $\sigma=25\%$ & \$1.5052 & \$3.6467 & +\textit{142\%}\\\hline\hline
Age = 67 & $\sigma=8\%$ & \$0.0470 & \$0.0939 & +\textit{100\%}\\\hline\hline
& $\sigma=15\%$ & \$0.3230 & \$0.7043 & +\textit{118\%}\\\hline\hline
& $\sigma=20\%$ & \$0.6362 & \$1.4534 & +\textit{128\%}\\\hline\hline
& $\sigma=25\%$ & \$1.0060 & \$2.4051 & +\textit{139\%}\\\hline\hline
\end{tabular}
\end{center}

The above table illustrates the impact of investment (RPI) volatility
($\sigma$) on both the RCLA and S-RCLA value, assuming the same Gompertz
mortality with parameters $\lambda=0$, $m=86.3$ and $b=9.5$. Note that both
the RCLA and S-RCLA are represented per guaranteed dollar of lifetime income
(i.e. scaled) and that the valuation rate (and hence $\mu$) is equal to 5\%.
We assume no deferral ($\tau=0$) and hence no bonus ($\beta=0$). The table also
displays the percent by which the Super RCLA exceeds the RCLA value, under
various volatility assumptions and ages.
\end{table}

Table \#4\textbf{ }illustrates the impact of investment (RPI) volatility
($\sigma$) on both the RCLA and S-RCLA value, assuming the same Gompertz
mortality with parameters $m=86.3$ and $b=9.5$. Note that both the RCLA and
S-RCLA are represented per guaranteed dollar of lifetime income and the
valuation rate (and hence $\mu$) is equal to 5\%. We assume no deferral ($\tau
=0$) and hence no bonus ($\beta=0$). The table also displays the percent by
which the Super RCLA exceeds the RCLA value, under various volatility
assumptions and ages. Thus, for example, at the age of 67, under both a
valuation rate $\rho=5\%$ and a spending percentage $\gamma=5\%$, the value of
an S-RCLA is between 100\% and 140\% greater than the value of a basic RCLA,
depending on the level of volatility assumed in the RPI. It seems that under
greater volatility $\sigma$, not only are the values of RCLA and S-RCLA
higher, but the ratio between S-RCLA and RCLA is greater as well.

\subsection{Connection to Guaranteed Living Withdrawal Benefit (GLWB)}

As we alluded to in the introduction, variants of RCLA derivatives are
embedded within variable annuity (VA) contracts with guaranteed living income
benefits (GLiBs) sold in the U.S., with variants sold in the UK, Japan, and
now in Canada. This is now a market with close to \$1 trillion in assets, and
with annual sales of over \$100 billion, in 2008. Hence the motivation for
studying these products. A GLiB is a broad term that captures a wide variety
of annuity riders, including the Guaranteed Minimum Withdrawal Benefit (GMWB),
the Guaranteed Lifetime Withdrawal Benefit (GLWB) and the Guaranteed Minimum
Income Benefit (GMIB). Thus, for example, a typical GLWB assures the
policyholder that if they withdraw no more than \$5 per \$100 of initial
investment deposit, they will be entitled to receive these \$5 payments for
the rest of their life regardless of the performance of the investments. They
can withdraw or surrender the policy and receive the entire account value --
net of withdrawals to date -- at any time. On the other hand, if the account
value ever hits zero the guarantee begins and the annuitant receives lifetime payments.

Although in general the valuation of exotic options within retirement benefits
has been analyzed by Sherris (1995) for example, these more specialized GLWB
products have been studied by Dai, Kwok and Zong (2008) as well as Chen,
Vetzal and Forsyth (2008) and Milevsky and Salisbury (2006). Our paper
provides yet another perspective on these types of embedded options and Table
\#5 can now be interpreted as more than just model values for a theoretical
product, but an actual estimate of the discounted value of the embedded
insurance offered by a variable annuity with a guaranteed lifetime withdrawal benefit.%

\begin{table}
\textbf{Table \#5a}

\begin{center}%
\begin{tabular}
[c]{||c||c||c||c||c||}\hline\hline
\multicolumn{5}{||c||}{\textbf{Super GLWB Value with Deferrals \& Bonus: Medium
Volatility (}$\sigma=17\%$)}\\\hline\hline
\multicolumn{5}{||c||}{\$100 investment into a Guaranteed Lifetime Withdrawal
Benefit (GLWB)}\\\hline\hline
Initial Purchase & Bonus, Deferral, Spending & $\rho=3.0\%$ & $\rho=5.0\%$ &
$\rho=7.0\%$\\\hline\hline
Age = 50 & $\beta=5\%,\tau=1,\gamma=5\%$ & \$39.5199 & \$19.0804 &
\$8.2223\\\hline\hline
& $\beta=5\%,\tau=7,\gamma=5\%$ & \$40.3168 & \$18.4768 &
\$7.5435\\\hline\hline
& $\beta=5\%,\tau=10,\gamma=5\%$ & \$38.8829 & \$17.0176 &
\$6.6352\\\hline\hline
& $\beta=5\%,\tau=15,\gamma=5\%$ & \$34.6250 & \$13.7056 &
\$4.8320\\\hline\hline
& $\beta=5\%,\tau=20,\gamma=5\%$ & \$28.5642 & \$9.9539 &
\$3.0714\\\hline\hline
Age = 65 & $\beta=5\%,\tau=1,\gamma=5\%$ & \$13.0509 & \$6.1738 &
\$2.5882\\\hline\hline
& $\beta=5\%,\tau=7,\gamma=5\%$ & \$10.8703 & \$4.7075 &
\$1.8006\\\hline\hline
& $\beta=5\%,\tau=10,\gamma=5\%$ & \$9.0896 & \$3.6539 &
\$1.2920\\\hline\hline
& $\beta=5\%,\tau=15,\gamma=5\%$ & \$5.9436 & \$2.0457 &
\$0.6090\\\hline\hline
& $\beta=5\%,\tau=20,\gamma=5\%$ & \$3.1648 & \$0.9087 &
\$0.2177\\\hline\hline
Age = 70 & $\beta=5\%,\tau=1,\gamma=5\%$ & \$7.1354 & \$3.3118 &
\$1.3634\\\hline\hline
& $\beta=5\%,\tau=7,\gamma=5\%$ & \$5.2707 & \$2.1989 & \$0.8097\\\hline\hline
& $\beta=5\%,\tau=10,\gamma=5\%$ & \$4.0458 & \$1.5467 &
\$0.5182\\\hline\hline
& $\beta=5\%,\tau=15,\gamma=5\%$ & \$2.1754 & \$0.6977 &
\$0.1911\\\hline\hline
& $\beta=5\%,\tau=20,\gamma=5\%$ & \$0.8564 & \$0.2258 &
\$0.0487\\\hline\hline
Age = 75 & $\beta=5\%,\tau=1,\gamma=5\%$ & \$3.2413 & \$1.4654 &
\$0.5891\\\hline\hline
& $\beta=5\%,\tau=7,\gamma=5\%$ & \$2.0320 & \$0.8077 & \$0.2834\\\hline\hline
& $\beta=5\%,\tau=10,\gamma=5\%$ & \$1.3834 & \$0.4970 &
\$0.1559\\\hline\hline
& $\beta=5\%,\tau=15,\gamma=5\%$ & \$0.5568 & \$0.1647 &
\$0.0411\\\hline\hline
& $\beta=5\%,\tau=20,\gamma=5\%$ & \$0.1368 & \$0.0329 &
\$0.0064\\\hline\hline
\end{tabular}
\end{center}

Notes: The table displays the value of a CONTINUOUS step-up Guaranteed
Lifetime Withdrawal Benefit (GLWB) under a variety of bonus, deferral and
withdrawal assumptions. It is the value of the Super-RCLA multiplied by the
number of lifetime dollars guaranteed. The mortality is assumed Gompertz with
parameters $\lambda=0$, $m=86.3$ and $b=9.5$. Prices are risk neutral (ie.
$\mu=\rho=r=$ risk-free rate).
\end{table}

\begin{table}
\textbf{Table \#5b}

\begin{center}%
\begin{tabular}
[c]{||c||c||c||c||c||}\hline\hline
\multicolumn{5}{||c||}{\textbf{Super GLWB Value with Deferrals \& Bonus: Low
Volatility (}$\sigma=10\%$)}\\\hline\hline
\multicolumn{5}{||c||}{\$100 investment into a Guaranteed Lifetime Withdrawal
Benefit (GLWB)}\\\hline\hline
Initial Purchase & Bonus, Deferral, Spending & $\rho=3.0\%$ & $\rho=5.0\%$ &
$\rho=7.0\%$\\\hline\hline
Age = 50 & $\beta=5\%,\tau=1,\gamma=5\%$ & \$22.8628 & \$6.9956 &
\$1.3465\\\hline\hline
& $\beta=5\%,\tau=7,\gamma=5\%$ & \$22.6176 & \$6.1226 &
\$1.0295\\\hline\hline
& $\beta=5\%,\tau=10,\gamma=5\%$ & \$21.8057 & \$5.3677 &
\$0.8103\\\hline\hline
& $\beta=5\%,\tau=15,\gamma=5\%$ & \$19.5396 & \$3.9733 &
\$0.4759\\\hline\hline
& $\beta=5\%,\tau=20,\gamma=5\%$ & \$16.1957 & \$2.6430 &
\$0.2351\\\hline\hline
Age = 65 & $\beta=5\%,\tau=1,\gamma=5\%$ & \$5.4466 & \$1.4094 &
\$0.2317\\\hline\hline
& $\beta=5\%,\tau=7,\gamma=5\%$ & \$4.2610 & \$0.9040 & \$0.1189\\\hline\hline
& $\beta=5\%,\tau=10,\gamma=5\%$ & \$3.5343 & \$0.6506 &
\$0.0719\\\hline\hline
& $\beta=5\%,\tau=15,\gamma=5\%$ & \$2.2689 & \$0.3214 &
\$0.0249\\\hline\hline
& $\beta=5\%,\tau=20,\gamma=5\%$ & \$1.1458 & \$0.1229 &
\$0.0064\\\hline\hline
Age = 70 & $\beta=5\%,\tau=1,\gamma=5\%$ & \$2.4404 & \$0.5805 &
\$0.0887\\\hline\hline
& $\beta=5\%,\tau=7,\gamma=5\%$ & \$1.6694 & \$0.3148 & \$0.0369\\\hline\hline
& $\beta=5\%,\tau=10,\gamma=5\%$ & \$1.2655 & \$0.2038 &
\$0.0196\\\hline\hline
& $\beta=5\%,\tau=15,\gamma=5\%$ & \$0.6525 & \$0.0793 &
\$0.0052\\\hline\hline
& $\beta=5\%,\tau=20,\gamma=5\%$ & \$0.2322 & \$0.0209 &
\$0.0009\\\hline\hline
Age = 75 & $\beta=5\%,\tau=1,\gamma=5\%$ & \$0.8428 & \$0.1813 &
\$0.0254\\\hline\hline
& $\beta=5\%,\tau=7,\gamma=5\%$ & \$0.4822 & \$0.0793 & \$0.0082\\\hline\hline
& $\beta=5\%,\tau=10,\gamma=5\%$ & \$0.3215 & \$0.0446 &
\$0.0037\\\hline\hline
& $\beta=5\%,\tau=15,\gamma=5\%$ & \$0.1190 & \$0.0122 &
\$0.0007\\\hline\hline
& $\beta=5\%,\tau=20,\gamma=5\%$ & \$0.0247 & \$0.0018 &
\$0.0000\\\hline\hline
\end{tabular}
\end{center}

Notes: The table displays the value of a CONTINUOUS step-up Guaranteed
Lifetime Withdrawal Benefit (GLWB) under a variety of bonus, deferral and
withdrawal assumptions. It is the value of the Super-RCLA multiplied by the
number of lifetime dollars guaranteed. The mortality is assumed Gompertz with
parameters $\lambda=0$, $m=86.3$ and $b=9.5$. Prices are risk neutral (ie.
$\mu=\rho=r=$ risk-free rate). This Table \#5b is based on the RPI allocated
to low volatility investments.
\end{table}

Table \#5a\textbf{ }displays the value of a continuous step-up (a.k.a. super)
Guaranteed Lifetime Withdrawal Benefit (GLWB) under a variety of bonus, deferral
and withdrawal assumptions. We assume precisely the maximum permitted
withdrawals after the specified deferral, and no lapsation. Thus it is the value
of the S-RCLA multiplied by the number of lifetime dollars guaranteed based on
an initial deposit of \$100 The mortality is assumed Gompertz with parameters
$m=86.3$ and $b=9.5$. In contrast, Table \#5b\textbf{ }displays the same
\textquotedblleft super\textquotedblright\ GLWB, but under a low volatility of
$\sigma=10\%$. As in Table \#4b, the GLWB value is obtained by multiplying the
value of a S-RCLA by the initial number of dollars guaranteed.

So, for example, assume that a 65 year old deposits \$100 into a VA+GLWB that
offers a 5\% bonus for each year that withdrawals are not made, and it offers
a \textquotedblleft5\% of base'' payment for life once the income begins.
\ The underlying base -- on which the lifetime income guarantee is based --
steps up in continuous time. So, if the individual intends on holding the
VA+GLWB for 7 years, and then begins withdrawals, the value of this guaranteed
income stream (in addition to the market value of the account itself) is
$\$10.8703$ per \$100 initial deposit, under a 3\% valuation rate and
$\$4.7075$ under a 5\% valuation rate. This assumes the underlying VA assets
are invested in a portfolio of stocks and bonds with expected volatility of
$\sigma=17\%$. Again, note the contrast in GLWB values under a lower
investment volatility of $\sigma=10\%$ in Table \#5b. The same two benefits at
age $x=65$ are valued substantially lower at $\$4.2610$ under $\rho=3\%$ and
$\$0.9040$ under $\rho=5\%$.

This number comes from multiplying the S-RCLA value times five, since the
initial guaranteed amount is \$5. Of course, for there to be no arbitrage, the
ongoing management fees charged on the initial deposit of \$100 would have to
cover the discounted (time zero) value of the GLWB option. Once again, the
continuously stepped-up GLWB guarantee on a variable annuity policy is just a
bundle of S-RCLA units plus a portfolio of managed money in a systematic
withdrawal plan. As one would expect, the greater the volatility, the lower
the valuation rate and the younger the individual, the higher is the value of
the embedded option, at time zero.

\section{Conclusion and Discussion}

This paper values a type of exotic option that we christened a ruin-contingent
life annuity (RCLA). The generic RCLA pays \$1 per year for life, like a
classical deferred annuity, but it begins making these payment only once a
reference portfolio index is ruined. If this underlying reference index never
hits zero, the income never starts. The rationale for buying an RCLA, and
especially for a retiree without a Defined Benefit (DB) pension plan, is that
it jointly hedges against financial market risk and personal longevity risk,
which is cheaper than buying security against both individually. The
motivation for studying the RCLA is that this exotic option is now embedded in
approximately \$800 billion worth of U.S. variable annuity policies. The
impetus for creating stand-alone RCLA products is that they might appeal to
the many soon-to-be-retired baby boomers who (i.) are not interested in paying
for the entire variable annuity package, and (ii.) would be willing to
consider annuitization, but only as a worst case \textquotedblleft Plan
B\textquotedblright\ scenario for financing retirement. Indeed, there is a
substantial amount of economic and behavioral evidence -- see for example the
introduction to the book by Brown, Mitchell, Poterba and Warshawsky (2001) --
that voluntary annuitization is unpopular as a \textquotedblleft Plan
A\textquotedblright\ for retirees. Thus, perhaps a cheaper annuity, and one
that has a built-in deferral period might appeal to the growing masses of
retirees without Defined Benefit (DB) pension plans. This was suggested
recently by Webb, Gong and Sun (2007) as well, and has received attention
from both practitioners and regulators -- see for example Festa (2012). 

Our analysis is done in the classical Black-Scholes-Merton framework of
complete markets and fully diversifiable market (via hedging) and longevity
(via the law of large numbers) risk. We derived the PDE and relevant boundary
conditions satisfied by the RCLA and some variants of the basic RCLA. We then
described and used efficient numerical techniques to provide extensive
estimates and display sensitivities to parameter values.

Our simple valuation framework only provides a very rough intuitive sense of
what these ruin-contingent life annuities might cost in real life. Of course,
until a liquid and two-way market develops for these products, it is hard to
gauge precisely what they will cost in a competitive market. We are currently
working on extending the PDE formulation approach -- by increasing the number
of state variables in the problem -- to deal with stochastic mortality, which
might also be dependent on market returns, as well as the implications of time
varying volatility, non-trivial mortality risk, and mean reverting interest
rates. Likewise, we are investigating the game-theoretic implications of
paying RCLA premiums continuously, as opposed to up-front. In other words,
what happens when the RCLA option is purchased via installments, which then
endows the option holder (annuitant) to lapse and cease payment? What is the
ongoing No Arbitrage premium in this case? The option to lapse leads to a
variety of interesting financial economic questions regarding the existence of
equilibrium, all of which we leave for future research.%

As the U.S. Treasury and Department of Labor continues to encourage Defined Contribution (401k) plans to offer stand-alone longevity insurance to participants -- see for example the article by Lieber (2010) in the \emph{New York Times} -- we believe that research into the optimal design and pricing of these market contingent annuities will, in itself, experience much longevity.


\end{document}